\documentclass[11pt]{article}
\pdfoutput=1

\usepackage{graphics, color,soul}
\usepackage{graphicx}
\usepackage{amssymb}

\usepackage{lmodern,mathtools}

\usepackage{booktabs}
\usepackage[english]{babel}
\usepackage{amsmath,amssymb,amsbsy,amstext, amsthm, simplewick}
\usepackage{hyperref}
\usepackage{tikz}

\usetikzlibrary{decorations.pathmorphing}
\tikzset{snake it/.style={decorate, decoration=snake}}

\usepackage{amsfonts}
\usepackage{amssymb}
\usepackage{upgreek}
\usepackage{simplewick}
 \usepackage{exscale,relsize}
\usepackage{mathtools}
\usepackage{comment}

\usepackage[margin=1cm,labelfont={sf,bf,scriptsize},textfont={sf,scriptsize}]{caption}

\usepackage{colortbl}
\definecolor{lightgreen}{cmyk}{0.2, 0, 0.2, 0.2}
\definecolor{lightgray}{cmyk}{0.1,0.2,0,0.1}
\definecolor{lightgray2}{cmyk}{0.1,0.1,0,0.1}

\makeatletter
\newlength{\apb@width}
\newcommand{\autoparbox}[2][c]{\settowidth{\apb@width}{#2}\parbox[#1]{\apb@width}{#2}}

\makeatother

\numberwithin{equation}{section}

\def\beq{\begin{equation}}
\def\eeq{\end{equation}}

\def\bea{\begin{eqnarray}}
\def\eea{\end{eqnarray}}

\def\beq{\begin{equation}}
\def\eeq{\end{equation}}
\def\be{\begin{equation}}
\def\ee{\end{equation}}
\def\bea{\begin{eqnarray}}
\def\eea{\end{eqnarray}}

\def\0{{\vec{0}}}
\def\k{{\vec{k}}}

\def\x{{\vec{x}}}

\def\k{{\bf k}}

\def\x{{\bf x}}

\DeclareRobustCommand{\SkipTocEntry}[4]{}

\def\beq{\begin{equation}}
\def\eeq{\end{equation}}

\def\ba#1\ea{\begin{align}#1\end{align}}
\def\bg#1\eg{\begin{gather}#1\end{gather}}
\newcommand{\bseq}{\begin{subequations}}
\newcommand{\eseq}{\end{subequations}}

\DeclareSymbolFont{extraup}{U}{zavm}{m}{n}
\DeclareMathSymbol{\varheart}{\mathalpha}{extraup}{86}
\DeclareMathSymbol{\vardiamond}{\mathalpha}{extraup}{87}

\def\({\left(}
\def\){\right)}
\def\[{\left[}
\def\]{\right]}

\begin{document}

\begin{titlepage}

\setcounter{page}{1} \baselineskip=15.5pt \thispagestyle{empty}

\vbox{\baselineskip14pt
}
{~~~~~~~~~~~~~~~~~~~~~~~~~~~~~~~~~~~~
~~~~~~~~~~~~~~~~~~~~~~~~~~~~~~~~~~
~~~~~~~~~~~ }

\bigskip\

\vspace{2cm}
\begin{center}
{\fontsize{19}{36}\selectfont  \sc
TASI lectures on cosmological observables and string theory
}
\end{center}

\vspace{0.6cm}

\begin{center}
{\fontsize{13}{30}\selectfont  Eva Silverstein}
\end{center}


\begin{center}
\vskip 8pt

\textsl{
\emph{$^1$Stanford Institute for Theoretical Physics, Stanford University, Stanford, CA 94306}}


\vskip 7pt
\textsl{ \emph{$^2$SLAC National Accelerator Laboratory, 2575 Sand Hill, Menlo Park, CA 94025}}

\vskip 7pt
\textsl{ \emph{$^3$Kavli Institute for Particle Astrophysics and Cosmology, Stanford, CA 94025}}

\end{center}

\vspace{0.5cm}
\hrule \vspace{0.1cm}
{ \noindent \textbf{Abstract}These lectures provide an updated pedagogical treatment of the theoretical structure and phenomenology of some basic mechanisms for inflation, along with an overview of the structure of cosmological uplifts of holographic duality.   A full treatment of the problem requires `ultraviolet completion' because of the sensitivity of inflation to quantum gravity effects, including back reaction and non-adiabatic production of heavy degrees of freedom.    Cosmological observations imply accelerated expansion of the late universe, and provide increasingly precise constraints and discovery potential on the amplitude and shape of primordial tensor and scalar perturbations, and some of their correlation functions.     
Most backgrounds of string theory have positive potential energy, with a rich but still highly constrained landscape of solutions.  The theory contains novel mechanisms for inflation, some subject to significant observational tests.
Although the detailed ultraviolet completion is not accessible experimentally, some of these mechanisms directly stimulate a more systematic analysis of the space of low energy theories and signatures relevant for analysis of data, which is sensitive to physics orders of magnitude above the energy scale of inflation as a result of long time evolution (dangerous irrelevance) and the substantial amount of data.   {\it Portions of these lectures appeared previously in Les Houches 2013,  ``Post-Planck Cosmology'' .}
\vspace{0.3cm}
 \hrule

\vspace{0.6cm}}
\end{titlepage}

\tableofcontents






\section{Introduction and Motivations}

Inflation \cite{inflation} provides a mechanism for generating the structure in the observed universe in a very simple way: it is seeded by quantum fluctuations during a primordial epoch of accelerated expansion \cite{classicperts}.  This in itself is one of the most elegant results in physics, perhaps the most basic application of quantum field theory.  

Observational data sets\footnote{including the {\it Planck} 2015 release \cite{Planckpapers}\ and recent predecessors such as \cite{WMAP}\cite{SPT}\cite{ACT}, as well as ongoing searches for CMB B-modes \cite{Bmode}\ and large-scale structure studies} provide unprecedented results on the spectrum and statistics of primordial perturbations.  We will be particularly concerned with the tensor to scalar ratio $r$, the tilt $n_s$ and other features of the power spectrum, and non-gaussian corrections $f_{NL}^I$.  Other measurements such as direct gravity wave searches can also provide constraints and discovery potential, e.g. for cosmic strings.  This confluence of theory and observation is a good situation, even though both sides have significant limitations. 

The subject is far reaching in other ways, including nontrivial connections to string theory, which will be our focus in these lectures.  
Before getting to that let me dispel a possible misconception.  You will sometimes hear inflation described as fine tuned.  However, models of inflation -- including the simplest kind -- can easily be radiatively stable with dynamically generated couplings, i.e. fully `natural' from a Wilsonian low energy effective field theory point of view.  This is the case for some of the simplest classes such as chaotic inflation \cite{chaotic}\ or more general large-field inflation scenarios such as Natural Inflation \cite{Natural}, as well as some examples of small-field inflation such as \cite{smallsymm}.\footnote{A separate, very interesting question is how to think about initial conditions before inflation (or any alternative theory) -- there is no clear framework in which to discuss probabilities.  This is an interesting challenge, but does not affect the modeling and phenomenology of the perturbations any more than it affects other areas of physics.}  

Moreover, low energy field theory is all that is required to describe  most of the phenomenology;\footnote{Exceptions include contributions of defects such as strings, or situations where heavy sectors come in and out of the spectrum during the process.}   this goes back to pioneering works such as \cite{classicperts}\ for particular models, and one can capture the effect of enhanced symmetries on the physics of the perturbations in a very elegant, less model-dependent way using the recently developed effective theory \cite{EFT}\ or more general calculations such as \cite{generalsingle}.    However, without imposing particular symmetries, this theory leaves open an enormous space of potential possibilities for the observables.  The EFT Lagrangian even at the purely single-field level contains arbitrary functions of time.  Moreover, as we will see, the contributions of very heavy fields can make a non-negligible imprint on the observables if they couple significantly to the inflaton, requiring a multi-field description. 

It is also true, as we will discuss in detail, that the process would be strongly affected by the presence of Planck-suppressed higher dimension operators in the Lagrangian (there is sensitivity to an infinite sequence of such terms in the large-field cases just mentioned!).  This raises the question of their existence and robustness in a complete theory of quantum gravity.  In string theory we find interesting answers to these questions at the level of specific mechanisms for inflation.  Some of these in part realize the classic ideas, but they all bring substantial new twists to the story.  Examples include \cite{monodromy}\cite{KKLMMT}\cite{DBI}\cite{Trapped}\cite{roulette}\cite{unwinding}\ along with a number of other interesting proposals.   

Regardless of their fate as literal models of primordial inflation, novel string-theoretic mechanisms continue to contribute to a more systematic understanding of the process of inflation and its perturbations, leading in turn to a more complete analysis of cosmological data.  There is a rich synergy between ``top down" and ``bottom up" approaches.  

In these lectures, we will start with some basics of inflationary cosmology, early universe observables, and their sensitivity to high energy physics.  The energy scales involved in inflation are at most several orders of magnitude below the Planck scale, and the process can be described in the language of effective quantum field theory.  Nonetheless even Planck-suppressed quantum gravity corrections are {\it dangerously irrelevant} in the renormalization group sense, and cannot safely be neglected.   An important focus will be the most extreme version of this arising in large-field inflation, where we will cover the relation between primordial gravitational waves and quantum gravity and its symmetry structure.  The same is true of certain contributions of heavy fields with masses well above the Hubble scale of inflation, as has become clear recently \cite{production}.  As we will see in detail, this follows from the timescale involved in inflation, and also from the large (and growing) amount of relevant data.  

To set the stage, we will provide a brief, pedagogical introduction to the structure of string compactifications.  Upon reduction to four dimensions, these are massively dominated by cosmological backgrounds with positive scalar potential energy. Although there are (in)famously many solutions in the landscape, however, the structure is enormously constrained.  We will introduce several interesting mechanisms for inflation that arise consistently within these constraints.  Despite the richness of the landscape,  several of these dynamical mechanisms are sufficiently well-defined to motivate concrete empirical tests, which are underway.  Finally, we will describe how the structure of string compactifications -- particularly the absence of a hard cosmological constant, leading to their metastability-- fits in a highly nontrivial way with a direct method for uplifting the $AdS/CFT$ holographic duality to de Sitter spacetime.     

%



For lack of space and time, I will not be able to be comprehensive in these notes.  Some other interesting aspects of the subject are covered in Senatore's lectures at this school.  The literature contains numerous reviews such as \cite{inflationreviews}\cite{axionreview}\ and \cite{Dioreview}.

\section{Inflation: generalities}

The basic paradigm of inflationary cosmology requires a source of stress-energy $\rho$ in the Friedmann equations
\begin{equation}\label{Friedmann}
H^2 =\left(\frac{\dot a}{a}\right)^2 = \frac{\rho}{3 M_P^2}
\end{equation}
\begin{equation}
\frac{2 \ddot a}{a}+H^2=-\frac{p}{M_P^2}
\end{equation}
which dilutes slowly in the very early universe.\footnote{In these notes, we will not keep track of all the factors of order 1, focusing on the main physical points.}    Here $a(t)$ is the scale factor in an FRW metric
\begin{equation}
ds^2=-dt^2+a(t) d\vec x^2, ~~~ H=\frac{\dot a}{a},
\end{equation}
and for inflation we require 
\begin{equation}\label{Hslow}
\frac{\dot H}{H^2}, \frac{\ddot H}{H^3} \ll 1 
\end{equation}
for a period under which $a$ expands by a factor of roughly $e^{60}$.  After that, this source must decrease in favor of the radiation and matter domination in later epochs of cosmology.    

The early inflationary behavior occurs if the energy density driving inflation dilutes slowly, something which can arise in a wide variety of ways.  Because of the need for an exit into the later phase of decelerating FRW cosmology, we require the source $\rho$ to then decrease rapidly after inflation.  The potential energy $V(\phi)$ of a dynamical scalar field can play this role, and we will focus on this case.  

One broad class of mechanisms is known as slow roll inflation, where this potential energy $V(\phi)$ dilutes slowly because the potential is very flat, and then steepens (either in the $\phi$ direction itself or in some transverse direction), leading to the required exit.  Other mechanisms maintain a nearly constant potential energy even if $V(\phi)$ is steep, with interactions that slow the field down. 
These two classes of mechanisms lead to very different phenomenology, with the interacting theories leading to stronger -- and differently shaped-- non-Gaussian correlations among quantum fluctuations in the early universe.    

Given a single scalar inflaton, we can write an action of the form
\begin{equation}
\int d^4x \sqrt{-g}\left(\frac{M_P^2}{2}{ R}+{ L_{scalar}}(\phi, g^{\mu\nu}\partial_\mu\phi\partial_\nu\phi)\right)
\end{equation} 
as long as higher derivatives (acceleration) of the fields are not important in the solutions we consider.  If the potential energy $V(\phi)$ dominates over the kinetic energy in the solution, then the system undergoes accelerated expansion.  
To get an idea of the range of possibilities even at the single-field level, consider the following two classes of models: 

$\bullet$  Single-field slow roll inflation
\begin{equation}\label{SRaction}
S_{SR}=\int d^4 x \sqrt{-g}\left({\cal R}+(\partial\phi)^2-V(\phi)\right)
\end{equation}
with $V'(\phi)/(M_P V), V''/VM_P^2\ll1$.  These flatness conditions -- which can be radiatively stable as we review below -- ensure that the potential energy dominates for a sufficiently long time (roughly 60 e-foldings) to dilute early relics such as monopoles and curvature (or more conservatively, to at least persist through the window of $\sim 7$ e-foldings accessible to CMB observations).  

$\bullet$ DBI inflation \cite{DBI}\
\begin{equation}
S_{DBI}=\int d^4 x \sqrt{-g}\left({\cal R}-\frac{\phi^4}{\lambda}\sqrt{1-\lambda\frac{(\partial\phi)^2}{\phi^4}}-\Delta V(\phi)\right)
\end{equation}
with a steep potential.  In this case, the potential energy dilutes slowly because the interactions contained in the Born-Infeld action simply prevent the field from rolling fast,  $|\dot\phi|<\phi^2/\lambda$.  This mechanism was inspired by the DBI action arising on branes in highly redshifted string compactifications.  There, this constraint on the field velocity corresponds to causality:  the brane cannot move faster than the speed of light in the radial direction.  This notion of causality in field space is interesting, and is part of the way in which gravity emerges from the AdS/CFT correspondence.  

This example made it clear that single-field inflation has a much wider range than (\ref{SRaction}), with observational consequences in the level of `non-Gaussianity' (the size of the $N>2$-point correlation functions of primordial perturbations which we will review shortly).  As such, it helped to stimulate a more comprehensive field-theoretic treatment of inflationary observables \cite{EFT}.  

This is an example of how string theory participates in standard science, without being a framework that is practically falsifiable overall (as far as we know).  Some researchers argue that one should speak only of what is observable, essentially forgetting entirely about examples of UV completions or even semi-UV quantum field theoretic completions of the theory of the inflationary perturbations.  This is in some sense a principled view, but taken to an extreme it seems to me to be counterproductive.  In no physical system -- even those with a detailed, well-tested theoretical description -- do we practically measure more than a set of measure zero of the theory's predictions.  The full structure of the theory is nonetheless an important part of the science.  Even in cases where the data does not distinguish different theories, the existence of these mechanisms is worthwhile knowledge.  Moreover, again and again in this subject, the discovery of mechanisms (often within string theory) predates a systematic fully bottom-up treatment.  This is not a criticism of the bottom-up approach, which can systematize the subject beautifully given appropriate assumptions.   

The addition of the scalar field to model inflation and its exit has a very important additional consequence.  This new quantum field fluctuates according to the Heisenberg uncertainty principle, as does the metric.  This leads to seeds for structure that are generated via inflation, after the background solution dilutes pre-existing inhomogeneities.   Let us quickly review these perturbations, as they are central to the observational probes of inflationary physics.  There are many reviews of this; examples of readable references include \cite{Juanperts}\cite{Mukhanovbook}\cite{Dodelsonbook}.  We can parameterize the metric as
\begin{equation}
ds^2=-N^2dt^2+h_{ij}(dx^i+N^idt)(dx^j+N^jdt)
\end{equation} 
with
\begin{equation}
h_{ij}=a(t)^2\left[e^{2\zeta}\delta_{ij}+\gamma_{ij}\right]
\end{equation}
and $N, N^i$ the lapse and shift, non-dynamical modes of the metric that enforce constraints.
During inflation, we have $a(t)\approx e^{Ht}$.  We can use time reparameterization to shuffle
the scalar degree of freedom between fluctuations $\delta\phi$ of the inflaton and the scalar mode $\zeta$ in the metric, giving
the relation 
\begin{equation}
\zeta \sim \frac{H}{\dot\phi}\delta\phi
\end{equation}
That is,
\begin{equation}
\langle\zeta_{k_1}\zeta_{k_2}\rangle\equiv P_\zeta \delta({\bf k}+{\bf k'}) = \frac{H^2}{\dot\phi^2}\langle\delta\phi_{k_1}\delta\phi_{k_2}\rangle 
\end{equation}
There are in general higher-point correlation functions, otherwise known as non-Gaussianity,
\begin{equation}
\langle\zeta_{k_1}\dots\zeta_{k_N}\rangle.
\end{equation}
These are functions of the momenta $k$, subject to rotational and translational symmetries.\footnote{See P. Creminelli's lectures on this topic for a systematic introduction; we will discuss some aspects of these signatures below.}

They are given by in-in correlation functions,\cite{Juanperts}
\begin{equation}\label{inin}
\langle in|\overline{T}\left( e^{i\int_{-\infty(1+i\epsilon)}^t dt_1{\cal H}_{int}}\right)
\delta\phi_{k_1}(t)\dots\delta\phi_{k_N}(t)T \left(e^{-i\int_{-\infty(1-i\epsilon)}^t dt_2 {\cal H}_{int}}\right)
  |in\rangle \,,
\end{equation}
where the operators are evolved with the free Hamiltonian.\footnote{See Senatore's lectures for a pedagogical derivation.}

In single-field slow-roll inflation, as we will review shortly one finds $\langle\delta\phi_{k}\delta\phi_{k'}\rangle\sim \frac{H^2}{2 k^{3}}\delta({\bf k}+{\bf k'})$, leading to
\begin{equation}\label{powerspectrum}
\langle\zeta_{k}\zeta_{k'}\rangle\sim \frac{H^4}{2\dot\phi^2k^{3} }\left(\frac{k}{k_0}\right)^{n_s-1}\delta({\bf k}+{\bf k'}) ~~~~ {\rm (slow-roll)}
\end{equation}
where in this last formula we allowed for a nonzero tilt to the spectrum, parameterized by $n_s$ and computable in any specific model by keeping track of the leading effects of the slow variations of the fields.  For single-field slow-roll inflation, the non-Gaussianity is negligible.  More generally one must calculate the correlation functions using the interaction Lagrangian including all fields that contribute detectable effects \cite{Creminelli,ghost,DBI,generalsingle, production}, and they depend on the underlying model parameters.  A clean result that comes out of these calculations and the effective field theory treatment of the perturbations  \cite{EFT}\ at the purely single-field level is a relation between the sound speed of the perturbations and the level of non-Gaussianity, as well as theorems about the $k_i$-dependence (the shape) of the non-Gaussianity.   

Tensor modes $\gamma_{ij}$ also fluctuate; for polarizations $s_1,s_2$ one finds
\begin{equation}
\langle \gamma_{s_1,k_1}\gamma_{s_2,k_2}\rangle\equiv P_\gamma \delta_{s_1s_2}\delta({\bf k}+{\bf k'})\sim \frac{2 H^2}{M_P^2}\delta_{s_1s_2}\delta({\bf k}+{\bf k'})
\end{equation}
These produce a very important signal detectable indirectly through B-mode polarization in the CMB \cite{Bmode}.  This quantity is captured by the tensor to scalar ratio 
\begin{equation}
r= \frac{P_\gamma}{P_\zeta}.
\end{equation}  

The evolution and freezeout of modes in QFT on de Sitter spacetime is a special case of particle production in time dependent quantum field theory.    Consider a particle $\chi$ with a time-dependent frequency $\omega_k(t)$ for momentum $k$.    This time dependence could come from its coupling to the metric of an expanding universe, or from its coupling to other rolling fields (e.g. the inflation).   For such particles, we have the mode expansion in creation and annihilation operators
\begin{equation}\label{chigen}
\chi(t,\x)=\int_\k  a^{(in)}_{\chi, \k} v_k(t) e^{i\k\cdot \x}+h.c.\,,
\end{equation}
where the mode function $v_k(t)$ is a solution of the free equation of motion including the effects of the time-dependent frequency, 
\begin{equation}\label{vSchrod}
-\ddot v -\omega_\chi^2(t) v=0
\end{equation}
This is of the form of a Schrodinger problem, with effective potential $-\omega_\chi^2(t)$.    
Here the lowering operator is defined to satisfy $a^{(in)}_{\chi, \k}|in\rangle =0$, with $|in \rangle$ the `in vacuum' with no particles initially.   This encodes the evolution of the operator via the free Hamiltonian, appropriate for the interaction picture.  

When the evolution is nearly adiabatic, in between any particle production events at times $t_n$, a WKB approximation applies in which the solution is a linear combination of modes which we can write to good approximation as
\be\label{vsoln}
a^{3/2} v_k(t)=\alpha_{k }^{(n)} \frac{\exp(- i \int_{t_n}^{t} dt' \omega_\chi(t'))}{\sqrt{2\omega_\chi (t')}}
+{\beta_{k }^{(n)}}^* \frac{ \exp(i\int_{t_n}^{t} dt' \omega_\chi(t'))}{\sqrt{2\omega_\chi (t')}}, ~~~~ t_n+t_{pr}<t<t_{n+1}-t_{pr}
\ee   
with normalization $|\alpha_{k}^{(n)}|^2-|\beta_{k}^{(n)}|^2=1$, where $t_{pr}$ is the timescale of the production event during which $\dot\omega/\omega^2$ is most significant.   A simple example to keep in mind is one with 
\begin{equation}\label{tsquared}
\omega_\chi(t)^2=k^2+m_\chi(t)^2 = k^2+\mu^2+g^2\dot\phi^2 (t-t_n)^2
\end{equation}
with the mass minimal at $t=t_n$, where particle production is most efficient.  The notation here is natural in situations with a quartic $g^2\phi^2\chi^2$ coupling of $\chi$ to a rolling scalar $\phi$.  

The Schrodinger problem is then that of an inverted harmonic oscillator.
We can consider a mode solution which is pure positive frequency initially, 
i.e. take $\alpha_{k}^{(0)}=1, \beta_{k}^{(0)}=0$.  After the event,  a nontrivial $\beta$ contribution is generated generically, given the time dependence.  In the analogue Schrodinger problem, this is analogous to the presence of both an incident and reflected wave on one side of the potential barrier (with a single transmitted wave on the other side corresponding to the pure positive frequency initial condition in the time-dependent problem).   

The linear transformation between the solutions can be traded for the famous Bogoliubov transformation
$a\to \alpha a+\beta a^\dagger$.  The state $|in\rangle $ (which does not evolve under the free Hamiltonian)  now the squeezed state
\begin{equation}\label{squeezed}
|in\rangle={\cal N} \exp\left(\int\!\frac{d^3 \k}{(2\pi)^3}\frac{\beta_{k}}{2\alpha_{k}^*} a_{\k,out}^\dagger a_{-\k, out}^\dagger\right)|out\rangle\,,
\end{equation}
This solves the condition $(\alpha^* a_{out}+\beta a^\dagger_{out}) |in\rangle \equiv 0$, as can be seen by representing the lowering operator $a $ as $\partial/\partial a^\dagger$ or equivalently applying the commutator.  This trick is also useful for proving that the expectation value of the number operator is given by
\begin{equation}
\langle in| n_k |in \rangle =|\beta_k|^2
\end{equation}
There are several methods to solve the example (\ref{tsquared}), including analytic continuation at large complex $t$, a saddle point calculation \cite{backdraft}, or brute force via parabolic cylinder functions.  The result is
\begin{equation}\label{beta0}
|\beta_{k}|^2=\exp\left(-\frac{\pi(\mu^2+\k^2)}{g\dot\phi}\right)
\end{equation}
This leads to a number density of produced particles
\begin{equation}\label{nchi}
\langle n_\chi \rangle \equiv \bar n_\chi = \int \frac{d^3 \k}{(2\pi)^3 a_n^3} |\beta(k)|^2
\simeq (g\dot\phi)^{3/2}\exp\left(-\frac{\pi\mu^2}{g\dot\phi}\right)\,,
\end{equation}
It is exponentially suppressed in the minimal mass $\mu$, with the denominator in the exponent reflecting the time dependence. 


Finally, let us apply this general formalism to massless scalar perturbations in de Sitter spacetime, satisfying an equation of motion  
\begin{equation}
\delta\ddot\phi_k+\frac{k^2}{a(t)^2}\delta\phi_k+3 H \delta\dot\phi_k = 0
\end{equation}
At early times, the scale factor $a(t)=a_0e^{Ht}$ is small, and the first two terms dominate, giving familiar Minkoski free field modes.  At late times, the middle term becomes negligible and there is a constant solution. 
One can easily check that the mode solution corresponding to the vacuum at early times is
\begin{equation}
u_k(x)=\frac{H}{\sqrt{2}k^{3/2}}\left(i+\frac{k}{aH}\right)e^{i\frac{k}{aH}}e^{i\k\cdot\x}
\end{equation}
which exhibits the anticipated behavior.  The two point function is
\bea
\langle in |\delta\phi(x)\delta\phi(x')|in\rangle &=& \int \frac{d^3\k}{(2\pi)^3} u_k(x) u_k^*(x') \\
&=& \frac{H^2}{2} \int \frac{d^3\k}{(2\pi)^3} \frac{e^{i\k\cdot(\x-\x')}}{k^3}\left(1+(\frac{k}{aH})^2 \right).
\eea
From this, working at late times (large $a(t)$) and transforming to momentum space reproduces the result quoted above,
\begin{equation}
\langle in | \delta\phi_\k \delta\phi_{\k'}|in \rangle = \frac{H^2}{2 k^3}\delta^{(3)}(\k+\k')
\end{equation}
from which one obtains the leading approximation to the power spectrum of inflationary scalar perturbations (\ref{powerspectrum}).  

\subsection{Inflationary dynamics and high energy physics}

Now let us return to the question of the mechanism behind inflation.  There are many possibilities.  
Let us start with a classic example of slow-roll inflation with scalar field Lagrangian \cite{chaotic}
\begin{equation}\label{msquared}
L_{matter}=\frac{1}{2}(\partial\phi)^2-\frac{1}{2}m^2\phi^2.
\end{equation}
This has the minimal number of parameters required to describe inflation and the exit from inflation.
According results reported very recently in \cite{Keck14}, it is now strongly disfavored by current data.  As such, a new parameter has been discovered (if we declare the origin in parameter space to be this simple model).  As we will see, the direction it goes is consistent with a simple energetic argument \cite{flattening}:  any
additional fields, even heavy ones, can adjust in an energetically favorable way to lower the potential energy.

The Friedman equation and scalar equation of motion give
\begin{equation}
H^2=(\frac{\dot a}{a})^2=\frac{1}{2}\frac{m^2\phi^2}{3 M_P^2}+kinetic 
\end{equation}
\begin{equation}
\ddot\phi + 3H\dot\phi=-m^2\phi\\
\end{equation}
which lead to $\dot\phi\sim m M_P$.  
The condition that the potential energy dominates is  then
\begin{equation}
\dot\phi^2\ll V(\phi) \Rightarrow m^2 M_P^2\ll m^2\phi^2 \Rightarrow \phi \gg M_P
\end{equation}
and in general in this super-Planckian regime $\phi\gg M_P$, the slow roll conditions 
\begin{equation}\label{SR}
\frac{M_P V'}{V}\ll 1, ~~~ M_P^2 \frac{ V''}{V} \ll 1
\end{equation}
are very well satisfied.   The number of e-foldings of inflation goes like $\phi^2/M_P^2$, and to match the observed normalization of the power spectrum we require  $m\sim 10^{-6}M_P, H\sim 10^{-5}M_P$.  
Note that the scalar field ranges over a super-Planckian distance in field space, but  we are not reaching Planck scale energy densities -- that would be completely out of control.  A large field range is sensitive to quantum gravity effects, but in a more subtle way that we will discuss below.

Similar results hold for any potential which behaves as
\begin{equation}\label{phip}
V(\phi)\to\mu^{4-p}\phi^p ~~~ for ~~~  \phi\gg M_P
\end{equation}  
for any $p$ (integer or not).  Note that at large field values, there is no reason to consider only integer powers $p$, since Taylor expanding about the origin is not useful.  Of course a priori it would not be clear that such a simple form as (\ref{phip}) for the potential pertains at all over a large range of $\phi$ given unknown quantum gravity contributions to the effective action. 
One of the mechanisms that we will find in string theory is a similar potential with $p<2$ over a large field range \cite{monodromy}  (more generally potentials flatter than $m^2\phi^2$ at large field range \cite{flattening}, although there are also potential generalizations with $p<n$ starting from higher integer powers $n$ \cite{powers, recentmonodromy, DDST}).  There is a nice mathematical structure behind such solutions which we will discuss. 

Before getting to that, let me first note that this model -- and any with a shift symmetry broken mildly by a flat potential -- is radiatively stable against loop corrections \cite{smolin}.  This makes it technically natural, and it is fully Wilsonian `natural' if the small scale $\mu\ll M_P$ required for phenomenology is obtained dynamically as we will discuss further below.  This is analogous to other familiar examples, such as `natural' solutions of the electroweak hierarchy problem.  As in that case, it remains to be seen if nature is Wilsonian-natural, but from a basic effective field theory point of view inflation is not generally fine tuned.\footnote{This is important for studies of potential signatures of earlier epochs -- sometimes these are motivated by the claim that minimizing tuning requires restricting to the fewest e-foldings consistent with solving the horizon problem.   Although that argument is not correct, the possibility of observing physics from the onset of inflation is certainly an interesting possibility worth pursuing (particularly given the various low-$\ell$ anomalies in the current data).}    

The basic reason that we will find this $V\propto \phi^{p<2}$ behavior in some robust regimes is rather trivial \cite{flattening}; it can be seen by simply coupling $\phi$ to  additional massive degrees of freedom.  Let us dress up the model (\ref{msquared}) by coupling the inflaton $\phi$ to a heavy field $\chi$ via canonical kinetic terms and
\begin{equation}\label{flattoy}
V(\phi,\chi)= \frac{1}{2}\phi^2\chi^2 + \frac{1}{2}M_\chi^2(\chi-m)^2
\end{equation}   
Here we take $M_\chi$ large compared to other scales in the problem.  The second term by itself would favor $\chi=m$, which when plugged into the first term would produce simply an $\frac{1}{2}m^2\phi^2$ potential, the theory (\ref{msquared}).  Instead what happens is that as $\phi$ builds up potential energy, $\chi$ shifts away from its minimum in an energetically favorable way, leading to a potential flatter than $\frac{1}{2}m^2\phi^2$.  
Classically integrating out $\chi$ -- i.e. solving its equation of motion $\partial_\chi V=0$ (it turns out a good approximation to neglect its kinetic term) -- leads to an effective potential for $\phi$ which is totally flat at large field range (i.e. $p=0$),
\begin{equation}\label{flatV}
V(\phi,\chi_*(\phi))=M_\chi^2 m^2\frac{\frac{1}{2}\phi^2}{\frac{1}{2}\phi^2+M_\chi^2}.
\end{equation}
In general, the ultraviolet completion of gravity might come with massive degrees of freedom -- certainly this is true of string theory, our leading candidate.  Although (\ref{flattoy}) is just a toy model, this effect occurs in string theory, leading in the large-field case to examples with potentials of the form (\ref{phip}) with $p<2$.  

Let us call this the flattening effect, for lack of a better term (it need not always produce a power law potential).
In the most recent models constructed in string theory, the inflaton potential provides one of several terms responsible for stabilizing other `moduli' fields.    As one example, fluxes in string compactifications 
(see (\ref{Ftildesquared})) below) contribute to the moduli potential.  In general, these generalized fluxes $\tilde F$ contain direct dependence on axion fields, which if slow rolling preserve the role of this term in moduli stabilization.  This effect can operate in many contexts -- not just axion inflation -- and provides a simple explanation for why all forms of inflation in string theory (which comes with many heavy fields that can adjust) remains observationally favored compared to the minimal scenario (\ref{msquared}).

Another classic example of inflation from the bottom up is known as Natural Inflation \cite{Natural}, with an axion potential of the form
\begin{equation}\label{Natural}
V(\phi) = V_0+ \Lambda^4 sin(\phi/f)
\end{equation}  
in terms of the canonically normalized scalar field $\phi =f \theta$ (where $\theta$ has period $2\pi$).  

There is an elegant quantum field theory motivation for this structure in terms of couplings of Yang-Mills fields to pseudo-scalars.\footnote{See e.g. \cite{Coleman}\ for a pedagogical introduction to some of the background.}   
It arises from quantum effects of a non-abelian Yang-Mills theory to which the axion couples only via its derivative.    There is a term in the action of the form $\int \theta Tr \epsilon_{\mu\nu\sigma\rho} F^{\mu\nu} F^{\sigma\rho}$ where $F$ is the field strength of a non-abelian Yang-Mills theory analogous to QCD and $\epsilon_{\mu\nu\sigma\rho}$ is a totally antisymmetric tensor.  For constant $\theta$, this term is a total derivative, integrating to zero, for topologically trivial configurations of the Yang-Mills fields.  Topologically nontrivial instanton configurations that contribute to the Feynman path integral of the theory have an integer-quantized value of $\int Tr  \epsilon_{\mu\nu\sigma\rho} F^{\mu\nu} F^{\sigma\rho}$, so such contributions are periodic in $\theta$, hence the sinusoidal potential (\ref{Natural}).  These contributions can also naturally be small in magnitude; instanton effects scale like $e^{-8\pi^2/g_{YM}^2}$, exponentially suppressed for small Yang-Mills coupling $g_{YM}$.    

Since taking derivatives of this potential with respect to $\phi$ brings down powers of $f$, it is straightforward to show that inflation on this potential requires $f > M_P$, and again the field rolls over a super-Planckian range.  

So far we have discussed two bottom-up examples of large-field inflation.  
From the point of view of low energy field theory, one can consider other models where the field rolls over a smaller range, including examples with inflection points, or hybrid inflation \cite{hybrid}\ with a second field that develops an instability that triggers an exit from inflation.  

Having warmed up with these examples, let us elaborate on the question of how to package the effects of ultraviolet degrees of freedom\footnote{As is common in high energy theory, we use `ultraviolet' to mean high energy here, in analogy to the high frequency part of the electromagnetic spectrum just beyond the visible range.} with a lightning review of effective field theory.  We can organize the Lagrangian of for example a weakly interacting scalar field theory in terms of a sequence of operators with different scalings under dilatations of space and time. 
The action is dimensionless -- it is the phase in the Feynman path integral.  The kinetic term $S_{kin}=\int d^4 x\sqrt{-g}\frac{1}{2}(\partial\phi)^2$ implies that a canonically normalized scalar field has dimension 1: the action is invariant under a scaling $x^\mu\to \eta x^\mu,\phi\to \phi/\eta$.    
Writing the action for $\phi$, taylor expanded about some point $\phi_0$ in field space and also expanded in derivatives, we have 
\begin{eqnarray}\label{Wilsac}
S=S_{kin}-\int d^4 x \sqrt{-g}\{\frac{1}{2}m^2(\phi &-&\phi_0)^2+\lambda_1(\phi-\phi_0)+\lambda_3(\phi-\phi_0)^3 +\lambda_4(\phi-\phi_0)^4 \nonumber \\
 &+& \lambda_6\frac{(\phi-\phi_0)^6}{M_*^2}+\lambda_{4,4}\frac{(\partial\phi)^4}{M_*^4}+\dots \}
\end{eqnarray} 
With the scalar mass $m$, and mass scale $M_*$ included as indicated, all terms in the action are dimensionless (with a convention that the coefficients $\lambda$ are dimensionless).  
In a given interaction term, let us call the fields and derivatives ${O}$.\footnote{${ O}$ stands for ``operator", an abuse of language in the Lagrangian formalism.}  Under the rescaling just described, ${ O}\to \frac{1}{\eta^\Delta}{O}$ where here $\Delta$ is determined by the above scaling of $\phi$ and its derivatives (in a strongly interacting theory these values are not accurate but here we can expand about weak coupling).   By dimensional analysis, the strength of the intereaction -- the effective coupling  -- generated by a given term is of order
\begin{equation}
\lambda_{eff}=\lambda\left(\frac{E}{M_*}\right)^{4-\Delta}    
\end{equation}
For operators of dimension $\Delta<4$, their effects grow at low energies, whereas for most operators, $\Delta>4$ and they are irrelevant at low energies.  

However, there is a subtlety here that is important for inflation: operators can be ``dangerously irrelevant".  Even if we stick to low energy densities, well below the Planck scale, higher-dimension operators actually matter for inflation, even if suppressed by $M_*=M_P$.  Adding a dimension-6 operator $V (\phi-\phi_0)^2/M_P^2$ would shift the slow roll parameters (\ref{SR}) by order 1, whereas they must be much smaller for inflation.  

In a Wilsonian view of field theory, if there is a high energy scale $M_*$ in the problem (such as an inverse lattice spacing in condensed matter physics, or the Planck scale $M_P$ in gravity) we would generically expect it to contribute a full sequence of higher dimension operators as in (\ref{Wilsac}) that are consistent with the symmetries or approximate symmetries of the model.  Even if we started with an action containing none of the $\Delta>4$ terms, we would generate them upon path integrating over high energy degrees of freedom.  
In the simplest case, this is a classical effect, as in the above discussion of ``flattening" in the model (\ref{flattoy}).  Its potential  (\ref{flatV}), obtained by solving $\chi$'s equation of motion (classically integrating it out), has an infinite sequence of terms suppressed by powers of the heavy mass scale $M_\chi$.  

The caveat about symmetries is very important:  if we do not start with large symmetry breaking, then
integrating over high energy degrees of freedom does not generate such terms.  As mentioned above, the relevant symmetry in inflation is the approximate shift symmetry $\phi\to\phi+const$.  If that is weakly broken by a very flat potential, quantum effects do not ruin it.  In the case of large-field inflation which is sensitive to an infinite sequence of $M_P$-suppressed operators, the question becomes one of whether the full theory has such a symmetry.    

In string theory the answer is affirmative, along appropriate directions in field space.\footnote{These directions are a good fraction of the scalar field directions in the theory; we will discuss some basic aspects of the spectrum and interactions of string theory below.} 
The mechanism that comes out -- a structure like a wind-up toy known as monodromy -- is a kind of hybrid of chaotic inflation and Natural inflation, but with new elements.    The essential features can be understood as in the following setup.  For one version of the mechanism, this turns out to be a dual, effective description of a local piece of the internal dimensions in the presence of the axion field.  The various ``branes", the extended objects of string theory, can end on each other in various ways.   Strings end on D-branes, and more generally there is a zoo of consistent configurations like this involving the higher dimensional objects.

\begin{figure}[htbp]
\begin{center}
\includegraphics[width=5cm]{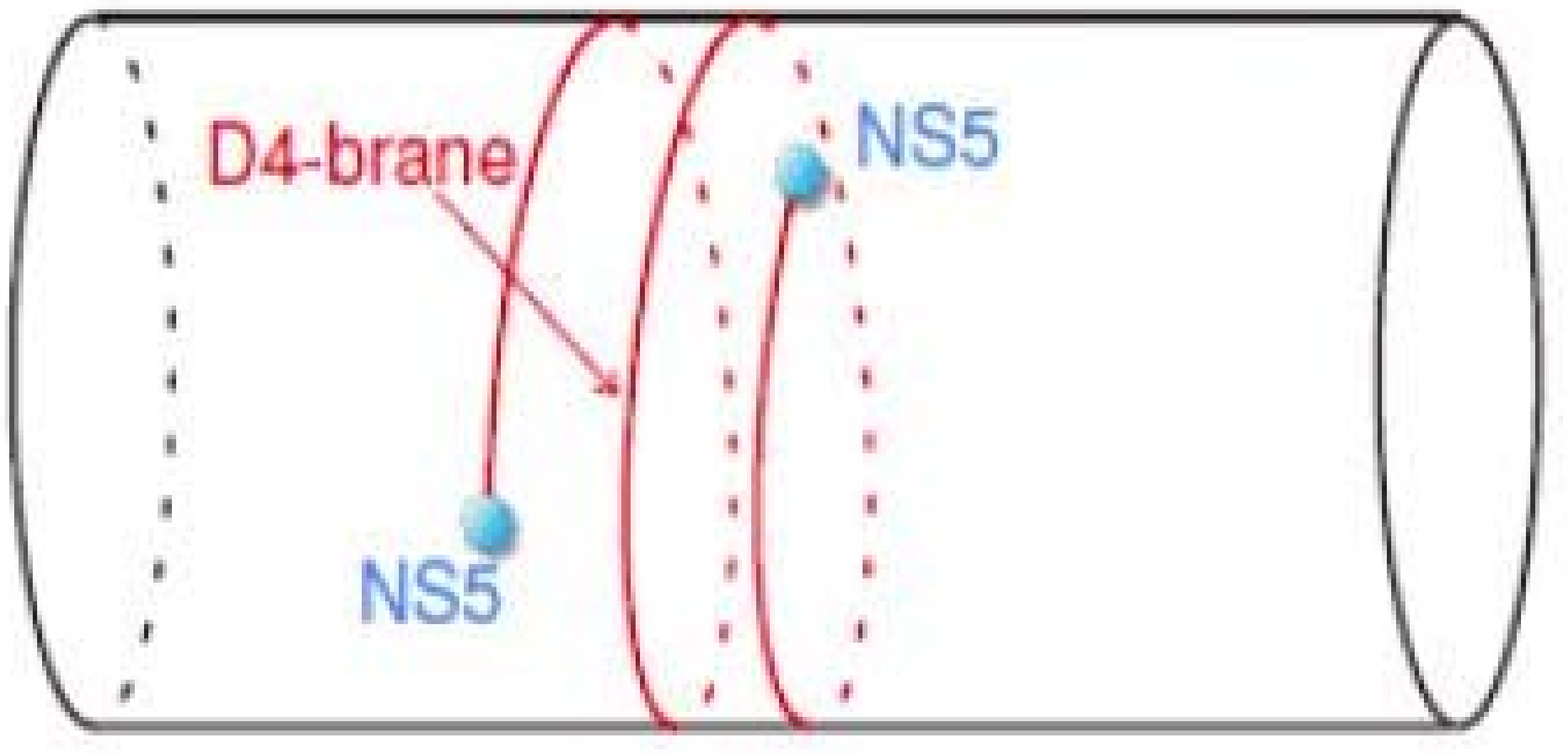}
\end{center}
\caption{}
\label{winduptoy}
\end{figure}


 In figure \ref{winduptoy}, we have two types of branes, one of which ends on the others.  These fill all four ordinary dimensions, and are situated as drawn on a space which is locally a cylinder within the extra dimensions.  These objects -- in particular the endpoints -- can move, and each such collective coordinate behaves as a scalar field in four dimensions.      
Before we include the stretched branes, moving one of the others around the cylinder constitutes a periodic direction. Let us call this direction $\theta$, with our scalar field $\phi$ being proportional to the distance $\ell_\theta\theta$.  However, once we include the stretched branes,  the scalar is no longer periodic:  the physics (in particular the potential energy built up by the stretched branes) depends on how many times we move around the underlying circle.  Specifically, using the fact that the energy density of the branes is given by their tension $\tau_4$  times the internal volume they wrap, we see that the potential energy will have the form
\begin{equation}
V(\phi) = \tau_4\sqrt{\ell_\perp^2 +\ell_\theta^2 \theta^2 }\equiv \mu^4\sqrt{1+\frac{\phi^2}{M_*^2}}
\end{equation}  
where the scale $M_*$ is determined by canonically normalizing the kinetic term for $\phi$.  This setup is rather simple and exhibits a potential of the form (\ref{phip}), with as in (\ref{flattoy}) a quadratic potential near the origin which flattens out at large field range.  Using the AdS/CFT description of the branes in terms of gravitational and matter fields (which we will discuss in the final lecture), this can be understood very explicitly as an example of the flattening effect just discussed \cite{flattening}.   

To belabor this rather trivially, let us replace $\theta$ in the formula above by $\theta-2\pi n$ for integer $n$.  This corresponds to starting in a sector with an $n$-times wrapped brane, and then winding it up or down further with the continuous variable $\theta$.  In this trivial sense, the whole system remains invariant under  $\theta\to\theta+2\pi n$, with a flattened large-field potential arising for any choice of $n$.

\subsection{Field Range and Tensor Mode Signature}

A very simple and important relation generalizes the above analysis of (\ref{msquared}) \cite{Lyth}.   
We can write the number of e-foldings as
\begin{equation}
N_e = \int \frac{da}{a}=\int \frac{\dot a}{a} dt = \int \frac{H}{\dot\phi}d\phi  = \int \frac{H M_P }{\dot\phi}\frac{d\phi}{M_P} = \sqrt{8} \,  r^{-1/2}\frac{\Delta\phi}{M_P}
\end{equation}
where in the last step we used the normalized results for the scalar and tensor power in single-field slow-roll inflation (even though in these lectures we are not keeping track of all `order 1' factors).  

This connection is extremely interesting, relating the excursion of the field in Planck units to an observable quantity.  Recall our discussion of Wilsonian effective field theory (\ref{Wilsac}).   A value $\Delta\phi>M_P$ corresponds to a model which is formally sensitive to an infinite sequence of higher dimension operators suppressed by $M_*=M_P$.  (If there are lower scales $M_*$ in high energy physics, the sensitivity is even greater.)  
When we plug in the numbers, a Planck range of field corresponds to $r\sim .002$ to $.01$ depending on the reheating temperature.
This level of $r$ is accessible observationally in the relatively near term according to the latest estimates \cite{Snowmass}.  It is exciting that relatively near-term observations will cover of order 10 Planck units in field range.


\subsection{Current Data: Constraints and Sensitivities}\label{data}

Let us briefly collect some of the central observables and summarize their current status.  

{\bf The cosmological constant}

Around 1998, several observations came together to establish the presence of accelerated expansion in the late universe. 
At least so far, this requires a single additional parameter (the cosmological constant $\Lambda$ of the standard $\Lambda$CDM cosmological model).   According to current data, time variations of the dark energy, as well as spatial curvature, are consistent with zero.  Although just a single parameter, this discovery has enormous implications.  

The causal structure of spacetime 
drastically changes upon changing the sign of the cosmological constant.  There are strong technical and conceptual reasons to suspect that the vacuum energy eventually decays in each observer patch, as we will discuss below.  We will see that almost all sources of four-dimensional potential energy in string theory are positive, and also that the theory contains the ingredients required to interpret the small value of $\Lambda>0$ as a selection effect \cite{Weinberg, BP}.  This does not imply that a selection effect is the correct interpretation, but currently it is the only viable proposal and is in some sense conservative.     

The cosmological constant raises enormously interesting questions, even if its value is a selection effect.  Among these are the interpretation of the de Sitter entropy, and generally the framework for observables in cosmology.  We will turn to these issues in the final section of these notes.  

{\bf B mode searches}

The tensor to scalar ratio $r$ is another single number with a great deal of theoretical significance.  It can most directly be constrained or detected via B-mode polarization in the microwave background, with the leading constraint as of this writing found in \cite{Keck14}.    There is currently a major worldwide effort to ultimately constrain $r$ to within roughly $\delta r< .001$.  This includes the ultimate ground-based program known as CMBS4, as well as existing ongoing programs \cite{Bmode}.    The observations of the coming years will be extremely powerful, covering a swath of $\sim 10 M_P$ in field range, and ultimately the field aims to decide the question of large versus small-field inflation.        

{\bf The scalar power spectrum and non-Gaussianities}

Scalar perturbations exhibit a nonzero tilt at $\sim 5\sigma$, consistent with the inflationary picture that the vacuum energy decreases slowly during the process.  The power spectrum and the higher point correlation functions are {\it a priori} rather general functions.  

On the one hand, the fact that to very good approximation the power spectrum can be described with only two parameters (its amplitude and tilt) strongly supports the inflationary paradigm, which would not generate wildly varying functions of scale.    
On the other hand, even within inflation, the detailed shapes and amplitudes of the $N$-spectra are not fixed by low energy considerations (such as \cite{EFT}, which contains arbitrary functions of time even at the single-field level, and which must be supplemented by any relevant additional fields, even when they are very heavy \cite{production}).  

As described in more detail in Senatore's lectures in this volume, inflation is perfectly consistent with some level of non-Gaussianity, and the current constraints on certain shapes it could take are consistent with zero.  They have not reached the threshold deciding between slow roll or non-slow roll forms of inflation, but have cut down on the parameter space of the latter.  The evolution of the inflaton along its trajectory in field space can easily be perturbed in various ways by periodic features in the potential and/or couplings to other fields.  Constraints on this can provide interesting bounds on such sectors, and there remains nontrivial discovery/constraint potential along these lines.

The signal/noise ratio for an $N$-point correlation function (with the noise here being the intrinsic quantum noise of the primordial fluctuations) is given by 
\begin{equation}\label{SNN}
(S/N)_N=\frac{\langle \zeta^N\rangle}{\sqrt{\langle\zeta^N \zeta^N\rangle}}
\end{equation}
In many cases, the denominator is well approximated by the disconnected component $\langle\zeta^2\rangle^{N/2}$. 

This is to be compared to the basic estimate for the sensitivity of the data:
\begin{equation}\label{pixels}
S/N\sim \frac{1}{\sqrt{N_{pixel}}}\sim 10^{-3}
\end{equation}
where the last expression is for the {\it Planck} CMB map.  The number of accessible modes will increase with large scale structure studies as well as 21 centimeter observations; the specific forecasts for this seem to be a subject of current debate in astrophysics.    

This sensitivity (\ref{pixels}) makes it possible to search for small effects of heavy degrees of freedom involved in UV completion.  Any new analysis requires a shape that is orthogonal, in the appropriate noise-weighted sense \cite{shapes}, from previously searched templates.  We will summarize a current example below.

\section{String Theory as a UV Completion:  effective action, stress energy sources, and symmetries}

This sensitivity to Planck suppressed operators is ample motivation to investigate inflationary cosmology in string theory.  Of course we do not know if string
theory is the UV completion of gravity in our world, but it is a very strong candidate to play this role.  Extensive work strongly suggests that the theory is internally consistent, satisfying numerous mathematical ``null tests", including concrete calculations of black hole entropy (albeit for highly supersymmetric examples).  In what follows we will study inflationary cosmology in the framework of string theory.  This leads to a number of different mechanisms for inflation, some observationally testable. 

 With this multiplicity of possibilities, it is not possible (at least with our current understanding) to falsify string theory as a whole; one can only test particular mechanisms.  But that is how physical model building often works.  Within the framework of quantum mechanics or quantum field theory, one needs to specify the system under study (the content and Hamiltonian of the model) in order to define and implement experimental probes.  The novelty here is that inflationary cosmology strictly speaking must be modeled in a UV complete theory, so we work in string theory as a candidate.   But otherwise it is similar -- within that framework we must specify the model under discussion (or at least a class thereof) to set up empirical tests in the cases for which that is possible.  

Of course, the inverse problem is very difficult -- given the data available, it is not in general possible to hone in aritrarily closely to a specific model.  But the data that is expected to become available can make broad distinctions such as how $\Delta\phi$ compares to the Planck range, and whether multiple fields or nontrivial interactions are involved in inflation.  
Data on non-Gaussianity can make a distinction between single field slow roll and other mechanisms for inflation.  And some models produce detailed signatures involving structure in the power spectrum or non-Gaussianity.  
In any case, regardless of the fate of particular models, string theory at the very least plays a useful role in suggesting a variety of novel mechanisms for inflation which feed into the bottom up treatment of signatures and data analysis, helping to make it more systematic.  Plus it is intellectually interesting!      

So without further ado, let us develop what we will need of string theory.  The effective action in $D$ dimensions is of the form 
\begin{equation}\label{Dac}
S =\frac{1}{2\alpha'^{\frac{D-2}{2}}}\int  d^Dx\sqrt{-G} \,e^{-2 \phi_{s}} \left(R-\frac{D-10}{\alpha'}+ 4 (\partial \phi_{s})^2 \right) + S_{{matter}}\,.
\end{equation}
for backgrounds with a sufficiently large curvature radius and small string coupling $e^{\phi_s}$, where $S_{matter}$ contains various matter sources that we will discuss in some detail below.  (See also \cite{TASIlectures, landscapereviews}\ for lecture notes devoted to the relevant features of string compactifications.)

The term proportional to $D-10$ exhibits the special nature of $D=10$:  it is the total dimensionality in which the classical theory has no potential energy and hence admits a Minkowski space solution.  The cases $D\ne 10$ are not anomalous:  the Weyl anomaly you may have learned about is a sickness that arises on the string worldsheet if you incorrectly solve the spacetime equations of motion, looking for a Minkowski solution in the presence of the nontrivial potential for $\phi_s$.  

It turns out that another special feature of the theory in $D=10$ is that it has a lot of supersymmetry (if one does not include generic matter sources), leading to an exact flat space solution.  One can show that physical transitions connect these different cases \cite{simeon}\cite{DimMutation}, so the picture to have in mind is that full theory has different limits with different effective $D$.  There are interesting behaviors at large $D$, including a spectrum dominated by $\sim 2^D$ axion fields, and simplifications of interactions and the loop expansion \cite{LargeD}.   
   
In the real world, scalar potential energy plays an important role in early universe cosmology as well as in the electroweak theory, and supersymmetry has not been observed despite significant LHC searches (although it remains an interesting possibility).  So as far as I can tell, the common approach setting $D=10$ from the start may be missing the big picture.  It has been natural to construct mechanisms for de Sitter and dark energy in the case $D>10$ because of the leading contribution of the above scalar potential \cite{SCdS,SCDE}.  

Having made this point, let me emphasize that there are important mechanisms for producing de Sitter solutions in $D=10$ that have been much more extensively studied, especially \cite{KKLT}\ and also other examples such as \cite{large, Saltman}\ and recently \cite{Danielsson}\ (to name just a few).  From the bottom up it remains an important possibility, even if it requires these special choices.  There are interesting signatures of the case where supersymmetry is broken by the inflationary Hubble scale in the early universe \cite{BaumannGreenSUSY}.    

But this pattern -- that it requires very special choices to suppress positive potential energy -- persists in the remaining contributions to the action.  The next most relevant term is the Einstein term proportional to the scalar curvature.   Roughly speaking, upon dimensional reduction, positively curved internal spaces (such as the spheres and other Einstein spaces that figure into AdS/CFT examples) lead to negative potential energy, zero Ricci curvature (Calabi-Yau manifolds) give zero classical potential energy, and negatively curved internal spaces (including compact hyperpolic spaces and their products) give positive potential energy.   Once again, the latter is {\it by far} the most generic among these possibilities.   This is already clear in two dimensions, where the genus expansion is dominated by higher-genus Riemann surfaces (with the sphere and the torus -- i.e. a $2d$ Calabi-Yau -- being exceptional), and extends to higher dimensions (in $3d$ the famous Thurston classification illustrates this, with nearly every topology giving positive potential energy).  

It is worth emphasizing that with sufficiently strong warping (strong redshifts at some places in the compact dimensions), the calculation of the effective potential, incorporating the constraint equations of general relativity, is subtle and interesting \cite{DouglasPotential}.  In fact, one can eliminate one unstable direction in scalar field space by letting the scalar field evolve strongly in one direction, sourcing strong warping in that direction \cite{DDST}.      

Let us proceed to the other sources of potential energy.  The additional matter sources in $S_{matter}$ include higher dimensional analogues of electomagnetic fields and various types of localized defects.   These include strings, domain walls, and their higher dimensional analogues, as well as important but more exotic structures known as orientifolds.  We have schematically
\begin{eqnarray}\label{Smatter}
S_{matter} =\int d^Dx \sqrt{-G}\{ &-&\sum_{n_B}\tau_{n_B}\frac{\delta^{(D-1-n_B)}(x_\perp)}{\sqrt{G_\perp}}+ \sum_{n_O}\tau_{n_O}\frac{\delta^{(D-1-n_O)}(x_\perp)}{\sqrt{G_\perp} }\nonumber \\ 
&+& e^{-2\phi_s}|H_3|^2+\sum_p |\tilde F_p|^2+C.S.+h.d.\}
\end{eqnarray}  
Here $h.d.$ stands for higher derivative terms.  $C.S.$ stands for ``Chern-Simons" terms coupling defects to the gauge potential fields (analogues of the vector potential in electromagnetism) that they source, to be discussed further below.  

We will explain the rest of the notation as we go through each type of contribution in what follows.  To do so will take some time as we will build them up from more familiar physics.    

First, $n_B$ indexes branes of spatial dimension $n_B$ and tension $\tau_{n_B}$ localized at $x_\perp=0$ (we could use normal vectors to write this more covariantly).  This is just saying that the branes' contribution to the energy is given by the volume they wrap times their tension.  They also are charged under higher-dimensional generalizations of electromagnetic fields, a feature we will discuss extensively below.  Some of the branes, known as D-branes, have the property that strings can end on them rather than forming closed loops.  Their positions are dynamical, described by scalar fields living on the brane.  
   
Similarly $n_O$ indexes the exotic defects known as orientifolds, which carry negative tension.  They affect the global structure of the space -- in the simplest case an orientifold introduces a $Z_2$ identification on the coordinates transverse to it, and in more general cases such objects are associated with certain topological quantum numbers of the space.  As such, although they contribute negatively to the stress energy, they cannot be wantonly produced to lower the energy -- a good thing for the stability of the theory.  Their positions are not dynamical, in contrast to the branes.   

The terms in the last line are the kinetic and gradient terms for higher dimensional analogues of the electromagnetic fields $F_{\mu\nu}=\partial_\mu A_\nu-\partial_\nu A_\mu$.  We will be very interested in these fields, so it will
be useful to build up what we need starting from the familiar case of electromagnetism.  

In ordinary electromagnetism, the potential field $A_\mu$ is useful even though it provides a redundant description, with the physics invariant under transformations of the form $A_\mu\to A_\mu+\partial_\mu \Lambda$.  
The field strength $F$ is gauge invariant.
 Another important set of gauge-invariant physical quantities are Wilson lines $e^{i\int_\gamma A_\mu dx^\mu}$ around a closed path $\gamma$.  We will need to understand these because their higher dimensional generalization will lead to axion fields, so let us discuss it here while we are on the subject.  Let us consider the case where $\gamma$ goes around a non-contractible cycle, say we have a circle in one spatial direction $y$:
\begin{equation}\label{ymetric}
ds^2=-dt^2 + d\vec x^2 + L_y^2 dy^2, ~~~ y\equiv y+1.
\end{equation}
Define
\begin{equation}
a(x) =\oint dy A_y
\end{equation}      
Under a gauge transformation, $A_y\to A_y+\partial_y \Lambda$, so $a(x)$ is invariant since the circle has no boundary.  The field $a(x)$ is a scalar field in the remaining dimensions ($t,\vec x$).  It is given in terms of the vector potential $A$ as 
\begin{equation}\label{aform}
A_y=a(x)  ~~ i.e. ~~ A=a(x) dy \equiv a(x)\omega_1
\end{equation}
where $\omega_1=dy$ is a one-form.  

The kinetic energy term in the action for this field is  
\begin{equation}\label{akinetic}
\sim \int d^3 x L_y\dot a^2\times \frac{1}{L_y^2},
\end{equation}
obtained from the $\int \sqrt{-g}F_{0y}^2g^{00}g^{yy}$ term in the electromagnetic action 
$\int dt d\vec x dy\sqrt{-g} F_{\mu\nu}F^{\mu\nu}$, where $L_y$ is the radius of the circle (\ref{ymetric}).   Here the first factor of $L_y$ comes from integrating over the $y$ direction with length $\int dy\sqrt{g_{yy}} = L_y$, and the $1/L_y^2$ comes from the inverse metric factor $g^{yy}$ in the contraction of the field strengths.  Scalar fields analogous to this one are numerous in string theory.      

Next let us review some of the physics of charged particles interacting with the electromagnetic fields.  
A charged particle sources the electromagnetic field via a coupling directly to the potential field: the action includes a term
\begin{equation}\label{Acoup}
 \int d^4 x J^\mu A_\mu =\int_{worldline} A
\end{equation}
In varying the action with respect to $A$ to get its equation of motion, this contributes a source localized along the worldline of the charged particle. 
 
Quantum mechanically, this coupling encodes something else we know,  the Aharanov-Bohm effect:  the wavefunction of a charge particle develops a  phase $e^{i\oint A}$ if it circumnavigates a region with magnetic flux.  

Finally, we note another feature that will be important: magnetic fluxes on compact spaces are quantized:  $\int F = N$ is an integer (up to a numerical factor).  One way to see this is that it is required for consistency of the wavefunction of a charged particle that moves around a vanishingly small circle in the compact space:  the flux outside the circle had better be quantized so that the Aharanov-Bohm phase is trivial and the particle's wavefunction is invariant.  

Now we are ready to start generalizing this to string theory and explain the last line of (\ref{Smatter}).  
The most basic generalization of this structure is a potential field $B_{\mu\nu}$, antisymmetric in its indices, which couples to the two-dimensional string worldsheet in the same way as the previous example $A_\mu$ couples to a particle worldline:  the action has a term
\begin{equation}\label{Bworldsheet}
\int_{worldsheet}B = \int \frac{d^2\sigma}{\alpha'} \sqrt{\gamma}\epsilon^{ab}B_{MN}\partial_a X^M\partial_b X^N.
\end{equation}
where the string tension is denoted $1/\alpha'$.  
Here $X^M=X^M(\sigma^a)$ describe the embedding of the string in spacetime as a function of position $\sigma^a$ on the worldsheet, $\gamma_{ab}$ is the worldsheet metric and $\epsilon^{ab}$ is the totally antisymmetric tensor in two dimensions ($\epsilon^{01}=1=-\epsilon^{10}$). 
The corresponding field strength is (in analogy to $F=dA$)
\begin{equation}\label{Hflux}
H_3 = dB_2 ~~~ i.e. ~~~ H_{MNP} \propto \partial_{(M}B_{NP)} 
\end{equation}
where the subscripts indicate the rank of the field and $d$, the exterior derivative, means a totally antisymmetrized derivative (indicated by the parentheses in the last expression).  The kinetic term for $B$ is the term proportional to $H_3^2$ in our action (\ref{Smatter}) above.  The field strength $H_3$ is invariant under gauge symmetries
\begin{equation}\label{Bgauge}
B\to B+d\Lambda_1, ~~~ i.e. ~~~ B_{MN}\to B_{MN}+\partial_{(M}\Lambda_{N)}
\end{equation}
where again the parentheses indicate antisymmetrization. 
 
Analogously to (\ref{aform}), this  leads to scalar fields $b(x)$ from the configuration
\begin{equation}
B=b(x)\omega_2
\end{equation}
where $\omega_2$ is a two-form.  For example, consider a two-dimensional torus among the extra dimensions with coordinates $y_{1,2}$,
\begin{equation}
ds^2_{extra} = L_{y_1}^2dy_1^2 + L_{y_2}^2dy_2^2
\end{equation}
 this configuration is in a component description $B=B_{y_1y_2}dy_1\wedge dy_2=b(x)dy_1\wedge dy_2$.  The normalization is such that $b$ is dimensionless, and has an underlying period of $2\pi$ under which the stringy Wilson line $e^{i\int B}=e^{i b}$ is invariant.     
This two-form potential $B_{MN}$ is special in that it arises for all perturbative closed-string theories.   

Next, there is an immediate generalization of this potential field $B$ which couples to strings.  The theory also famously contains defects of other dimensions, the `branes'.  Similar remarks apply to higher-rank potential fields $C_{p-1}$, i.e. antisymmetric fields with $p-1$ indices, which are sourced by D-branes \cite{joebook}.  These again yield axion fields $c(x)=\int_{\Sigma_{p-1}} C_{p-1}$ where $\Sigma_{p-1}$ is a non-contractible $p-1$ dimensional subspace in the extra dimensions.  The field strength is $F_{p}=dC_{p-1}$.     

However, that is not the full story -- there is a very interesting interplay among these various fields and their gauge transformations which we are now ready to describe.  There are two aspects to this, one having to do with the generalization of the $F_{\mu\nu}F^{\mu\nu}$ terms in the effective action, and the other having to do with couplings between scalar axion fields and branes.  

You may have noticed in the above action (\ref{Smatter}) that we wrote $\tilde F_p^2$ rather than $F_p^2$.  The generalized field strengths $\tilde F_p$ are defined as follows:
\begin{equation}
\tilde F_p = F_p+B\wedge F_{p-2}=dC_{p-1}+B\wedge dC_{p-3}
\end{equation}  
These are invariant under the gauge transformation (\ref{Bgauge}) as long as we accompany it with the transformation $C_{p-1}\to C_{p-1}-\Lambda_1\wedge F_{p-2}$.  That is, the full transformation is
\begin{equation}
B\to B+d\Lambda_1, ~~~ C_{p-1}\to C_{p-1}-\Lambda_1\wedge F_{p-2}.
\end{equation}

As before, there is actually an analogue of this in ordinary physics:  upon spontaneous symmetry breaking of electromagnetism (as occurs for example in superconductors), the low energy effective action has the Stuckelberg term
\begin{equation}
(A+\partial\theta)^2
\end{equation}
where $\theta$ is the would-be Goldstone mode which shifts under the gauge symmetry:  $A\to A+d\Lambda, \theta\to\theta-\Lambda$.  This exhibits the gauge invariance of the underlying system in the phase with a massive vector field.  Our couplings 
\begin{equation}\label{Ftildesquared}
\tilde F^2 = ( F_p+B\wedge F_{p-2})^2
\end{equation}
are just higher-dimensional versions of this.  

When we dimensionally reduce the theory on a $D-4$ dimensional space $X$,\footnote{as a toy example you can keep in mind a manifold $X$ that has a two-torus factor} if there are nontrivial field strengths (fluxes) $F_{p-2}$, then there will be a classical potential for the $b(x)$ field we described above.  This is the generic situation, as turning off the fluxes requires making a very special choice.  This classical, non-periodic potential for $b$ is in contrast to traditional axions in quantum field theory, which develop a potential through non-perturbative effects in the Yang-Mills theory to which they couple, as discussed above (\ref{Natural}).  

One might na\"ively conclude from (\ref{Ftildesquared}) that we get a quadratic potential for $b(x)$.  This is true close to the origin $b(x) = 0$.  But when we turn on $b(x)$, then as we build up potential energy in this term, the heavy degrees of freedom (such as gradients of the fields) can adjust in a more energetically favorable way, as in the toy model of flattening discussed above (\ref{flattoy})(\ref{flatV}).  For example, the $B$ field can develop a spatial dependence that reduces its overlap with $F_{p-2}$, at the cost of turning on the term $H_3^2=dB^2$ since the gradients of $B$ that are contained in $H$ cost energy.  This is analogous to the sharing of potential energy between the mass term and quartic term in (\ref{flattoy}).  This will generically cause the potential to deviate from the na\"ive $m^2\phi^2$ behavior.  In some concrete examples, one can see explicitly how it flattens the potential \cite{flattening}.  

One such example has an equivalent description in terms of branes, something we will need to understand in its own right.  Consider the term (\ref{Bworldsheet}) coupling the string to the potential field $B_{MN}$ that it sources.  If the $B_{MN}$ field is constant ($H_3=dB=0$), then the term  (\ref{Bworldsheet}) is a total derivative:  it can be written as
\begin{equation}
\int\frac{ d^2\sigma}{\alpha'} \partial_a( B_{MN} \epsilon^{ab}\partial_a X^M\partial_b X^N)
\end{equation}  
If the worldsheet has a boundary -- as occurs in the presence of the D-branes on which open strings end,   
then this term does not automtatically integrate to zero.  In the presence of D-branes, there is a direct dependence of the effective action on $B_{MN}$, not just its field strength $H_3=dB$. 

The full worldsheet action in the Polyakov form depends on the spacetime metric $G_{MN}$ as well as on $B_{MN}$, in the combination $G+B$:
\begin{equation}\label{wsGB}
S_{worldsheet} = \int \frac{d^2\sigma}{\alpha'} \sqrt{\gamma}\left\{\gamma^{ab}G_{MN}\partial_a X^M\partial_b X^N+\epsilon^{ab}B_{MN}\partial_a X^M\partial_b X^N\right\}
\end{equation}
We would like to motivate this further, and also obtain the action on D-branes as a function of $G_{MN}$ and $B_{MN}$.

To move toward these goals, it proves useful to first consider the action for a relativistic particle, which is the Born-Infeld action
\begin{equation}\label{BI}
S_{particle}=-m\int dt \sqrt{1-\dot x^2}
\end{equation}
From this action, you can derive the standard equation of motion $dp/dt=0$, where  $p=m\gamma \dot x = m\dot x/\sqrt{1-\dot x^2}$.  This can equivalently be written as
\begin{equation}\label{particleBIagain}
S_{particle}=-m\int d{\xi^0} \sqrt{G_{MN}\partial_{\xi^0} X^M\partial_{\xi^0} X^N}
\end{equation}
where $\xi^0$ is the coordinate along the worldline of the particle.  The action (\ref{BI}) arises (in the flat Minkowski spacetime $G_{MN}=\eta_{MN}$ if we choose this coordinate such that $X^0=\xi^0$, the simplest choice.  In a more generic background geometry, the equation of motion from this action reproduces the geodesic equation for particle motion in general relativity.  

The action (\ref{BI}) arises from a particle analogue of (\ref{wsGB}).  Focusing on the metric dependence, we have
\begin{equation}\label{particlepoly}
\int d\sigma^0 \sqrt{g_{00}}\left\{ G_{MN} g^{00} \dot X^M\dot X^N +m^2\right\}
\end{equation}
Integrating out $g_{00}$, which means solving the worldline Hamiltonian constraint, produces (\ref{particlepoly}).  

The generalization of (\ref{particleBIagain}) to higher dimensions, with the $B$ field included, turns out to be the Dirac-Born-Infeld brane action
\begin{equation}\label{DBI}
S_{DBI}=-\tau_B\int d^d\xi\sqrt{det((G_{MN}+B_{MN})\partial_aX^M\partial_b X^N+f_{ab})}
\end{equation}
where $\tau_B$ is the brane tension.  Here the determinant is taken with respect to the indices $a,b$ which label directions along the worldvolume of the brane, and $f_{ab}$ is the field strength for an Abelian gauge group which lives on the brane.  As a check, in the case that the brane is just sitting there motionless, this gives an action which is just minus the tension $\tau_B$ times the volume wrapped by the brane.  (There is also a coupling analogous to (\ref{Acoup}) to the $C_p$ potential field sourced by the brane, but here let us focus on the $B$ and $G$ dependence.)  

It turns out that these two ways in which this direct dependence on $B$ arises -- from flux or D-branes,
are related by the AdS/CFT correspondence. There, as we will discuss further below, D-branes are described equivalently by a curved spacetime background with fluxes $F_q$ turned on. One of these descriptions is usually more useful than the other in a given regime.  

So far we have focused on the way the axion field $b(x)=\int B$ enters into the stress-energy sources, but it is also interesting to consider their dependence on axions that come from the other potential fields $C_{p-1}$.  From various dualities that connect different string theories, there are relations between the different axions $b(x)$ and $c(x)$.  Indeed, while $b(x)$ axions get a potential from D-branes, there are other types of branes which produce a potential directly for the $c(x)$ type axions.  To see this more directly, integrate by parts on the couplings $B_{MN} F_{i_1\dots i_{p-2}}F^{MN i_1\dots i_{p-2}}$ coming from the $\tilde F^2$ terms in the action.  This produces couplings of the schematic form $H F c$, indicating a potential for $c$ axions in the presence of generic fluxes just as we discussed above for $b$ axions.        

The axion fields $b(x),c(x)$, and the brane collective coordinates $X$ provide candidate inflaton fields in string theory, along with some of the other scalar field moduli.  We will discuss some examples below.  But first we should discuss more about the process of compactification to four dimensions, since there are other scalar fields (collectively called `moduli' $\phi_I$) and they must all be accounted for in looking for inflationary solutions.  

We are interested in the four dimensional effective action obtained upon dimensionally reducing the theory  (\ref{Dac})(\ref{Smatter}) down to four dimensions on some space $X$ of dimension $D-4$.  Deriving this in detail in general is difficult.  However it is possible to work in controlled limits where one can obtain it to good approximation.  In the context of inflation, note from our discussion below (\ref{Wilsac}) above that this requires controlling enough $M_*$ suppressed terms, where the high energy scales in the problem include the Planck scale $M_P$, but also lower scales such as Kaluza-Klein scales and the scale of the string tension.

To make things simple -- but not completely generic -- let us consider the case where the localized sources we consider, the branes and orientifolds, do not in themselves source strong warping (gravitational redshift) in most of the compact space.  This can be quantified as in \cite{micromanaging} (section 3.3) in terms of the amplituded of the internal gravitational potential in the solution.  More generally, one must analyze the problem as in \cite{Douglaswarping}.  Some of the following is contained in the lecture notes \cite{TASIlectures}\ as well as \cite{landscapereviews}\cite{Nil}\cite{Saltman}, so I will be relatively brief here and
focus on the main elements as well as some aspects that were understood more recently.    

The Einstein term $\int d^D x\sqrt{-G}R$  reduces to a the four-dimensional Einstein term, kinetic terms for the geometrical moduli fields, and a contribution to the potential energy $V(\phi_I)$.  It is useful to canonically normalize the kinetic terms.  Let us start with the four-dimensional Einstein term, in the original, `string', frame:
\begin{equation}\label{Rterm}
\int\frac{ d^4 x\sqrt{-g}}{\alpha'} (\frac{L^{D-4}}{g_s^2}+loops)\frac{ R}{2}.
\end{equation}
Here $L^{D-4}$ is the size of $X$ in units of the string tension $1/\alpha'$, and the ``+loops" includes quantum corrections to the Newton constant.
This look contribution can build up in the presence of a large number of species, and it in general will depend on the scalar field moduli since the effective cutoff scale will depend on them.   It will be interesting to analyze the effect of this term on moduli stabilization and inflation when it does dominate over the classical term, but for our purposes here we can for the most part ignore this correction.     

Since the size $L=e^\phi$ (in units of the string tension) and $g_s=e^{\phi_s}$ are dynamical scalar fields, it is most convenient to rescale the four-dimenisonal metric $g_{\mu\nu}\to (g_s^2/L^n)g_{\mu\nu}$ to work in Einstein frame in four dimensions.  Since then $\sqrt{-g}\to  (g_s^2/L^n)^2\sqrt{-g}$, this rescales the potential terms by  $(g_s^2/L^n)^2$,  a factor which strongly dilutes the potential energy at large radius and weak coupling.  In fact, all contributions vanish in that limit, something which has important implications both technically and conceptually as we will see.  

Notice that such a rescaling does not change the ratio of the Planck mass to the scale of the string tension $1/\alpha'$, which is
\begin{equation}\label{Planckmass}
M_P^2\alpha' \sim \frac{L^n}{g_s^2}
\end{equation}
When we go to Einstein frame, we work at a fixed value of $M_P$, with an Einstein term $\sim \int\sqrt{-g}M_P^2 R/2$.  

Next let us discuss the kinetic term for the overall volume mode.  We can write
\begin{equation}
ds^2=ds^2_{(1)}+e^{2\sigma} ds^2_{(2)}     
\end{equation}
with the first term the $4d$ metric with coordinates $x^\mu, \mu=0,1,2,3$ and the second that of the internal $k=D-4$ dimensions, with its overall volume pulled out as a scalar field $\sigma(x)$.  From this, we obtain for the scalar curvature
\begin{equation}
R=R_{(1)}+e^{-2\sigma} R_{(2)}-2k\nabla^2 \sigma-k(k+1) g^{\mu\nu}\partial_\mu\sigma\partial_\nu\sigma
\end{equation}
indicating that $\sigma$ is indeed proportional to the canonically normalized scalar field corresponding to the overall volume.  The precise formulas for the canonical volume and dilaton for arbitrary $D$ can be found for example in \cite{DDST}.

The internal curvature makes a positive contribution to the potential energy $V(\phi_I)$ if the internal space is negatively curved (for which there is an infinite number of possible topologies), and a negative contribution if it is positively curved; these contributons are of order $M_P^4(L^n/g_s^2)\times (1/L^2)\times (g_s^2/L^n)^2$.  Choosing a Ricci-flat internal space, a Calabi-Yau manifold, gives zero contribution to the potential at tree level; if one also chooses $D=10$ this preserves a nonzero fraction of the supersymmetry below the scale $1/L$.  Orientifolds  wrapping $n_O$ internal dimensions contribute negatively, $\sim -M_P^4 (L^n/g_s)\times (1/L^{n-n_O})\times (g_s^2/L^n)^2$. The fluxes $\tilde F_p^2$ with their legs along the internal dimensions contribute positively, a contribution going like $M_P^4(L^n)\times (\tilde N^2/L^{2p})\times (g_s^2/L^n)^2$.  (In all these contributions, the dependence on $g_s$ is not obvious but does come out of the string theory calculations \cite{joebook}.)

These various contributions individually contribute steep tadpoles, of the form
\begin{equation}
e^{\beta\phi_c/M_P}
\end{equation}
for the canonically normalized fields associated to the string coupling and internal volume.  However, we very recently found a string-theoretic version of assisted inflation \cite{assisted}\cite{TolleyWesley} -- where multiple exponential terms contribute in a way that reduces the $\beta$ coefficient along the rolling direction and stabilizes the transverse directions \cite{SCDE}.  This literal form of the potential does not produce observationally viable inflation, although it may be of interest for dark energy research depending on the structure of the Standard Model contribution; see also the recent work \cite{SandipDE}.  
These new solutions suggest the possibility of more economical mechanisms for treating moduli stabilization and inflation together.  More generally, the plethora of inflationary mechanisms that do not require single-field slow roll may help stabilize the moduli with fewer complications.   

That said, previous work has instead largely implemented inflation within a pre-existing scenario for stabilizing the string coupling, volume, and other moduli.  For the rest of these notes we will proceed that way, with a few additional comments below.       

Having built up an understanding of these basic ingredients, let us sketch the form of the potential.  Let us consider for illustration a compactification in type IIB string theory (in any dimensionality) with a single scale $L\sqrt{\alpha'}$, e.g. a product of Riemann surfaces with curvature radius of this order.  The internal volume is $\sim L^{D-4}$, we switch to Einstein frame as described above, and the various sources are diluted by appropriate factors.  We will describe the flux contributions in a little more detail shortly, but very schematically we have
\begin{eqnarray}
\frac{V}{M_P^4} &=& g_s^2\left[\frac{D-10}{L^{D-4}}-\frac{c_h}{L^{D-2}}+(\frac{N_3}{L^3})^2\frac{1}{L^{D-4}}+(NS ~ or ~ p-q ~ branes) \right] \\ 
&+& g_s^3\left[-\sum_O\frac{\hat \tau_O L^{n_O}}{L^{2(D-4)}} +\sum_D\frac{\hat \tau_D L^{n_D}\sqrt{1+b^2/L^{m_D}}}{L^{2(D-4)}}\right] \\
&+& g_s^4\left[(\frac{Q_3-C_0 N_3}{L^3})^2\frac{1}{L^{D-4}}+(\frac{Q_5-\frac{1}{2}c_2 N_3+\frac{1}{2}b N_3}{L^3})^2\frac{1}{L^{D-4}}+\dots
\right]\\
&+& h.d. + warping+quantum \\
\end{eqnarray}
Here the first line comes from $D$, the curvature, NS flux, and heavy branes.  The second line comes from orientifolds and D-branes, which can be localized along various submanifolds of different dimensionalities.
The third line comes from the generalized fluxes; we will shortly elaborate on the mechanism for flux stabilization of complex structure moduli.  Both the fluxes and branes can depend on axions, as also schematically indicated here. 

Since the potential drops to zero in the controlled regime of weak coupling and large radius, the simplest way to meta-stabilize these fields in expanding the potential in $g_s$ and $1/L$ is to play three terms off each other:  a leading positive term, a negative term to produce a dip in the potential, and a third positive term.  The $D-D_c$ term, negative curvature, $H$ flux, and certain branes can contribute the first term; orientifolds, positive curvature, and certain perturbative and non-perturbative quantum effects can contribute negatively; and the generalized fluxes $\tilde F$ naturally contribute the required positive third term.

Other fields, such as complex structure moduli of Riemann surfaces or Calabi-Yau manifolds, are stabilized in a relatively elegant way using fluxes.   The generalized flux terms can contain axion dependence (depending on the topology), but as we will see, out on a slow roll potential for such fields these generalized flux terms can still play their usual role in moduli stabilization, while supporting inflation.  For the next discussion I will suppress the axion dependence for brevity (and since its appearance depends on the topology and other features) but the simplification of having the inflation potential help stabilize moduli should be borne in mind.  

Flux stabilization works essentially as follows.  Intuitively, at fixed volume the flux on a given cycle prevents it from contracting since the energy density grows very large, and similarly the flux on its dual cycle prevents the latter from contracting.   More explicitly, the flux can be expanded in an integral basis of forms whose integrals over $a$ and $b$ cycles are given by
\begin{equation}
\int_{a_I} \alpha_J=\delta_{IJ}, ~~~~ \int_{b_I}\beta_J=\delta_{IJ}, ~~~~\int\alpha_I\wedge \beta_J=\delta_{IJ}
\end{equation}
In terms of these, we can write
\begin{equation}
F=\sum Q_I^{(a)}\alpha_I + Q_I^{(b)}\beta_I
\end{equation}
where the $Q$'s are the flux quanta.
We can also expand the field strength in terms of another basis in which the Hodge dual $*$ is simple:
\begin{equation}
\int_{a_I}\omega_J=\delta_{IJ}, ~~~ \int_{b_I}\omega_J=\tau_{IJ}, ~~~*\omega=-i\omega, ~~~ *\bar\omega=i\bar\omega 
\end{equation}
where $\tau_{IJ}$ depend on the complex structure moduli.  From this, and the linear relation between the two bases of forms, we obtain a potential energy for the complex structure moduli of the form (see e.g. \cite{GKP, Saltman}\ for more details).  
\begin{equation}
V(\tau)=\int F\wedge * F= Q^T A(\tau) Q
\end{equation}
where $A(\tau)$ is the matrix $i ((2\tau(\tau-\bar\tau)^{-1}\bar\tau, -(\tau+\bar\tau)(\tau-\bar\tau)^{-1}), (-(\tau-\bar\tau)^{-1}(\tau+\bar\tau), 2(\tau-\bar\tau)^{-1}))$ (see e.g. \cite{Saltman}).  One finds that with sufficiently general fluxes, this stabilizes the complex structure moduli.  

To sum up, roughly -- but very instructively -- speaking, the moduli are stabilized in one of two ways, via a 2 or 3 term structure in the potential.   For example, for the dilaton we have the form
\begin{equation}
V(g_s)=\tilde a g_s^2-\tilde b g_s^3+\tilde c g_s^4,
\end{equation}  
a metastable 3-term structure.  
For some ratios of cycle sizes, on the other hand, we have 2-term flux stabilization with the participation of axions, a simple example of flattening of the inflaton potential \cite{flattening, powers}.  Consider a product space with one component of volume ${\cal V}_1$ and another of volume ${\cal V}_2$, with the total volume ${\cal V}_1{\cal V}_2$ fixed.  With generalized fluxes $Q_1$ and $b Q_2$ on the two factors, we will get a potential for the ratio ${\cal V}_1/{\cal V}_2$ of the form
\begin{equation}
V(\frac{{\cal V}_1}{{\cal V}_2})=\frac{{\cal V}_1}{{\cal V}_2}Q_1^2+\frac{{\cal V}_2}{{\cal V}_1} (b Q_2)^2
\end{equation}
Solving for ${\cal V}_1/{\cal V}_2$, this potential is linear in the axion $b$ rather than quadratic.  This is a toy model of what happens in detail in more explicit models, such as \cite{flattening, powers, DDST}.    

Although for our simple discussion we focused on the case where the gravitational redshifts (warping) are not too strong, it is important to emphasize that significantly redshifted regions play a role in much interesting model-building, including  the KKLT construction \cite{KKLT}\ and inflationary models build from it.  The energy density of localized sources of potential energy described above gets redshifted in such regions, leading to naturally small scales.        

We will not spend more time on this subject, since there is an extensive literature including for example \cite{landscapereviews, TASIlectures, SCdS, KKLT, large, Saltman, Danielsson, Nil,nogo}.  
The reader may well wonder how the rich (but constrained!) mess that is the string landscape can lead to anything involving concrete observational predictions, beyond the general idea that some quantities such as the cosmological constant may consistently be interpreted as a selection effect \cite{BP}.  In particular, I have emphasized above some not-so-hidden assumptions involved in the focus on very specific classes of compactifications, choice of $D$, etc., and this would seem to exacerbate the naive problem.  
However, we will see in the next section that the structure of string compactifications lead to several concrete and reasonably generic inflationary mechanisms,  some of which make specific predictions.

\section{A sample of string-theoretic inflationary mechanisms and signatures}

Having developed some of the relevant background on scalar fields and the effective action in controlled regimes of string theory, we will now discuss several mechanisms with some interesting and robust features.  Many of the features of the string theory action and sources which we just explained will play a role in the following mechanisms, so we will refer back to the previous section as we go.    

By the way, most of the mechanisms described here were not found by seeking a string-theoretic generalization of bottom up models, or by looking to produce certain signatures.  Nor were they discovered by trying to be systematic from the bottom up.    Some were serendipitous discoveries obtained while studying other, only tangentially related, problems (such as those  we will discuss in the final section).  Such is the way it goes often in theoretical physics: it is not always very effective to try to legislate immediate progress in a particular direction, but it is useful to be aware of many problems in order to recognize potential solutions.      

\subsection{Axions in string theory, large-field inflation, and signatures}

Axion inflation and chaotic inflation are beautiful examples illustrating the Wilsonian naturalness of large-field inflation from the bottom up, while at the same time revealing important new effects that emerge in their closest analogues in string theory.\footnote{For a recent overview of various interesting versions of axion inflation with additional references, see \cite{axionreview}.}    

Axions in traditional quantum field theory couple derivatively at the classical level, as in the discussion below (\ref{Natural}).  In that model, known as Natural Inflation, there is a sinusoidal potential resulting from non-perturbative effects.  The model requires a super-Planckian period $f$ in order to inflate.  It is radiatively stable, protected by a shift symmetry, and has an appealing partial UV completion in terms of the Yang-Mills theory which generates the axion potential.    

\smallskip

A full UV completion requires quantum gravity to account for possible Planck-suppressed terms which could affect the results.  In the case of string theory, it turns out that axions work differently in two important respects:  

\begin{itemize}
\item The potential energy generically depends directly on the axion, although there is an underlying periodicity; this monodromy structure is as in the wind-up toy example above in figure \ref{winduptoy}.   

\item Even if we worked in a very special regime where we turned off the fluxes and branes which produce this direct dependence on the axions in the potential, the axion kinetic terms imply $f\ll M_P$ in controlled regimes (large radius and weak coupling) \cite{Banksaxion}.  There is, however, a multiple-field version of Natural inflation, known as N-flation \cite{Nflation}, which mitigates this via collective effects of multiple axion fields as we will explain below.

\end{itemize}

The first point follows from the couplings between $b,c$ and the fluxes and branes discussed in the previous section.  Let us consider, for example, the brane action (\ref{DBI}).  To be specific, consider for example a brane with five spatial dimensions (so $a=1=0,1,\dots,5$) wrapped on a two-dimensional torus $T^2$ of size $L\sqrt{\alpha'}$.  That means that along the torus directions, the metric $G_{MN}$ is $L^2$ times the identity, $L^2 (dy_1^2+dy_2^2)$.  The $B$ field is simply a $2\times 2$ antisymmetric matrix with entries $B_{12}=-B_{21}=b$. The worldvolume field strength is also antisymmetric, and its flux on the torus is quantized:  $\int_{T^2} f =2\pi N_3$.\footnote{The subscript ``3" here refers to the fact, which I haven't explained and we will not need directly, that $f$ on the brane sources the D3-brane potential field.}  

We can explicitly work out the determinant, working for simplicity with the coordinates $\xi^a=X^a$ (identifying the worldvolume coordinates with the corresponding embedding coordinates $X^M$ in spacetime).     If we put the brane at rest, the DBI action gives us a four-dimensional potential energy of the form
\begin{equation}\label{bPythagorean}
V=\mu^4\sqrt{L^4+(b+N_3)^2}
\end{equation}
Here the coefficient includes the Einstein frame conversion factor $(g_s^2/L^n)^2$ and other effects like gravitational redshift at the location of the brane.  As we mentioned above, we can work within a standard moduli-stabilization scenario and treat $g_s$ and $L$ as constant (for more details and nuances in specific examples, please see the original literature).    

This illustrates several things.  First, at large $b$ this in itself gives a linear potential.  Secondly, we see the monodromy (windup toy) structure as follows.  We are in a particular sector of flux with a given $N_3$:  it takes a non-perturbative process to change that quantum number.  We can move continuously in $b$ arbitrarily far.  If we move by a period in $b$, the same configuration could have been obtained starting from a shifted value of $N_3$.  So the whole theory -- the set of flux quantum numbers combined with $b$ field configurations -- respects the periodicity.  However, on a given branch, again, the $b$ field builds up more and more potential energy as it moves around the underlying circle multiple times.  In general there may be additional, residual effects of this underlying periodicity, leading to a sinusoidal contribution to the  potential as in (\ref{Vplusosc}) below.    

This structure of the axion potential is rather robust.  For a specific model realization, it must occur within a sufficiently stable compactification, with an explanation for the phenomenological value of $\mu^4$ (best thought about after canonically normalizing all fields using the results for the kinetic energy discussed below).  Originally this has been done in a modular way just to show it is consistent \cite{monodromy}:  stabilizing the dilaton and radii and other scalars using say \cite{KKLT}\cite{large}, on a manifold with gravitationally redshifted regions which naturally warp down the energy scales as in \cite{RS}\cite{GKP}.\footnote{The claim in \cite{conlon}\ regarding back reaction in these examples is incorrect. The author argues that since the charge of an object does not dilute when it is placed in a highly redshifted region, its energy cannot decrease either because of the BPS bound.  The question can be most simply addressed for electrons in a Schwarzchild black hole background.  The flaw in this reasoning is that the black hole itself is well above the BPS bound, and an electron near its horizon with redshifted energy $m_e\sqrt{g_{00}}$ does not cause any such violation of the BPS bound.  More generally, as we have seen above, back reaction can easily (and does in specific examples) flatten the potential.}  In particular, the Randall-Sundrum mechanism, realized in warped compactifications of string theory \cite{GKP}\cite{KKLT}, introduces scales which are exponentially suppressed as a function of the curvature radius of the geometry.  This means that the low scale of the coefficient in the potential is not tuned (again according to the Standard Wilsonian effective field theory); this is analogous to dynamical supersymmetry breaking as a theoretically natural explanation of the weak scale in particle physics.
More recently this mechanism has been realized in a more economical way, itself helping to stabilize the string coupling and radii \cite{SCDE}.  

Now let us next explain the second point about the decay constant $f$ being sub-Planckian.  This follows from the analogue of (\ref{akinetic}) for the $b$ and $c$ fields.  Note that the normalizations above are such that $b$ and $c$ are dimensionless variables, with the string-theoretic analogues of Wilson lines $e^{i \int B}=e^{i b}$ being invariant under a shift $b\to b+2\pi$.   

For the $b$ field, starting from the $\int \sqrt{-g}H_{0m_1m_2}g^{00}g^{m_1m_1'}g^{m_2m_2'}H_{0m_1'm_2'}/g_s^2$ term in the original action, we obtain a kinetic term for $b$ (or equivalently, the canonically normalized field $\phi_b=fb$)
\begin{equation}\label{bkinf}
\int \sqrt{-g}H^2\sim \int d^4 x\sqrt{-g} \frac{L^n}{g_s^2\alpha'}\dot b^2 \frac{1}{L^4}=\int d^4 x\sqrt{-g} f^2\dot b^2 \equiv \int d^4 x\sqrt{-g}\dot\phi_b^2
\end{equation} 
where the last factor of $1/L^4$ comes from the internal inverse metric factors $ g^{m_1m_1'}g^{m_2m_2'}$.  (For simplicity here we are taking all scales of order $L$; more generally one obtains similar results from a more precise calculation of the overlap of differential forms $\int \omega\wedge \star\omega$ that arises from plugging $B=b\omega$ into the $H^2$ term.)  
From this we read off
\begin{equation}
f_b\sim \frac{M_P}{L^2}
\end{equation}
This is much less than $M_P$ for the controlled regime $L\gg 1$.  
Similar results hold for the $c$ axions, which lead to a suppression with respect to the string coupling as well, for example for the axion coming from $C_2$ we get
\begin{equation}
f_{c_{2}}\sim\frac{g_s}{L^2} M_P \ll M_P.
\end{equation}
The upshot is that in controlled limits of string theory, we obtain sub-Planckian underlying axion periods.  

Combining this with the first bullet point above, the way axion inflation works in string theory is more like chaotic inflation, with a large field range and no periodicity along a given branch of the potential.  However, there is still an underlying periodicity in the setup, and that does lead to periodic corrections to the potential, with a model-dependent amplitude $\Lambda^4$.  Altogether, in terms of the canonically normalized field $\phi_{axion}$, the potential is of the form
\begin{equation}\label{Vplusosc}
V(\phi_{axion}) = V_0(\phi_{axion}) + \Lambda^4 \Upsilon(\frac{\phi_{axion}}{f})
\end{equation}  
where $V_0$ is derived from the brane or flux couplings as above, with special cases including a Pythagorean potential (\ref{bPythagorean}).   $\Upsilon$ is a periodic function, for example $sin((\phi_{axion}-\phi_0)/f$, that is generated by physics that is periodic under $b\to b+2\pi$.  This includes instanton effects, but can be more general.  In the windup toy picture figure \ref{winduptoy}, there is a nice mechanical way to see such effects.  As the two endpoints come together, new sectors of light strings connecting them appear, and their effects will in general generated an $\Upsilon$ term.      

In the windup toy picture of figure \ref{winduptoy}, the strength of the periodic term is determined by the lateral separation of the two endpoints on the cylinder.  In general, it depends on how the moduli are stabilized.  In some circumstances, this leads to a non-perturbative suppression of $\Lambda$, somewhat similar to what we discussed
in the original case of Natural Inflation where the amplitude $\Lambda$ was naturally an exponentially small instanton effect.  

One feature that deserves further analysis is to incorporate the moduli-dependence of the period $f$ into the dynamics, keeping track of the small adjustments that the moduli make even though they are heavy.  We discussed an analogue of this in the `flattening' analysis above, where such adjustments make an important difference to the shape of the potential. Adjustments of the heavy fields such as the moduli determining $f$ will cause the period to change at some level during the process of inflation.  This, along with the overall amplitude $\Lambda^4$, seems rather model-dependent, but it will be interesting to explore whether some general conclusions are possible about its direction or shape.    This has been explored in detail in \cite{drift}\ and the {\it Plank} inflation paper \cite{Planckpapers}.    

There is another feature of string theory as a UV completion which is worth remarking on:  there are in general multiple axions.  If a large number $N_a$ are light, they produce inflation over a smaller range of field for each axion \cite{assisted}\cite{Nflation}:  multiplying the kinetic term (e.g. (\ref{bkinf})) by $N_a$ increases the effective $f_{axion}$ by a factor of $\sqrt{N_a}$.     
On the other hand, they contribute to the renormalization of Newton's constant (\ref{Rterm}), so we get a range of the form
\begin{equation}
\frac{f^2}{M_P^2}\sim \frac{N_af_0^2}{M_{P,bare}^2+N_a\Lambda_c^2} 
\end{equation}
where $\Lambda_c$ is the appropriate cutoff scale.  So as emphasized in the original paper \cite{Nflation}\ this is not a parametric enhancement of $f$ at large $N$, but depending on the cutoff scale this can make a difference.  

Adding fields tends to push the tilt to the red side \cite{Liddlemulti}\ compared to the single-field version of the model, presumably because the field is rolling down the steeper part of each individual potential.  For a sinusoidal or effectively $m^2\phi^2$ potential along each axion direction, this moves the predictions further toward the outside of the allowed region in the data  (although one should not take $2\sigma$ distinctions seriously).  But as we have seen, that case is very special, more generically the potential is flattened as happens in axion monodromy inflation.
Putting all this together, perhaps a generic example would have $N_a>1$ light axions, each in a monodromy-expanded potential.  The phenomenology of this case is analyzed in \cite{Danjie}, with the expected redward shift of the tilt.  

\subsubsection{Additional fields and sensitivity to high mass scales}

A very recent development \cite{production}\ is the realization that exponentially suppressed non-adiabatic effects of very heavy fields can leave detectable/constrainable imprints on the CMB.  This can be understood by comparing (\ref{nchi}) and (\ref{pixels}), suggesting that masses even somewhat higher than the $\sqrt{\dot\phi}$ could be produced in a detectable way, since
\begin{equation}
\exp(-\pi \mu^2/g\dot\phi)\sim 1/\sqrt{N_{pixel}}
\end{equation} 
corresponds to a minimal mass $\mu\gtrsim (g\dot\phi)^{1/2}$.  This rough idea survives a detailed calculation in \cite{production}, which reveals several novel features of the resulting correlators.  Their shape is orthogonal to previously searched shapes, and the signal/noise in their $N$-point functions can be competitive with, or even greater than, their effects on the power spectrum.  This motivates a new data analysis, including in a regime where the effect grows mildly with $N$ and requires a novel search strategy.   With its discrete shift symmetry,  axion monodromy motivates such sectors of heavy fields, and provided part of the impetus for this new regime amenable to concrete observational tests. 

Finally, let us briefly describe two other examples that involve angular directions and multiple fields (in this case, light fields).  In one example, known as trapped inflation \cite{Trapped}, the field rolls slowly down a steep part of its potential, repeatedly dumping its kinetic energy into the production of particles (or higher dimensional defects) that become light along its trajectory in field space.  This is motivated by the quasiperiodic variables we have been discussing in string theory, where there can easily be a periodicity to particle production events.  To see this most easily, consider the regime of figure \ref{winduptoy}\ where the two endpoints are close together laterally.  In addition to the spacefilling brane sectors shown, the spectrum of the theory contains sectors of strings or branes stretched between the two endpoints.  If the two endpoints come close enough together each time around the underlying circle, these sectors can be non-adiabatically produced, with inflaton kinetic energy dissipated in the process, slowing down its motion down the potential.  This can produce inflation on a steep potential, with an equilateral non-Gaussian signature in the perturbations.  

Another multifield example is known as Roulette Inflation \cite{roulette}, where a pair of scalar fields organize into a modulus and phase $\rho e^{i\theta}$ related by (broken) supersymmetry in a particular scenario \cite{large}.  With this and other multifield examples including \cite{Nflation}, the results depend on the initial conditions, i.e. on the trajectory.  The analysis of the predictions is statistical, with ultimately an unknown distribution on the initial conditions.   The contributions of the additional fields to the fluctuations may lead to interesting large-scale anomalies and non-Gaussianity \cite{Dick}. 

\subsubsection{Phenomenology of axion inflation}

After fitting model parameters to the normalized power spectrum and $N_e$ (with some uncertainty in reheating dynamics), one obtains well defined predictions for $r$, $n_s$, and other quantities as a function of any remaining parameters in the model.  
The classic single-field quantum field theoretic axion theory, Natural Inflation, makes predictions along a swath of the $r-n_s$ plane which is indicated in figure 1 of the Planck inflation paper \cite{Planckpapers}.  

Axion monodromy inflation \cite{monodromy}\ makes distinct predictions.  It produces a detectable tensor signal within the range $.01 < r < .1$ which is observationally accessible but not yet constrained by the data.  Multiple B-mode experiments promise to cover this range and beyond 
\cite{Bmode}\cite{Snowmass}.  As discussed above, this is very exciting in general because it will cover the full range of possible large-field inflation with a super-Planckian field excursion $\Delta\phi$.
The Planck 2015 inflation paper \cite{Planckpapers}, and even more recently the BICEP/Keck 2014 data combined with {\it Planck} \cite{Keck14}, show the current $2\sigma$ limits on $r$ and the tilt $n_s$, including two representative versions of axion monodromy.  These constraints disfavor $\phi^p$ inflation with $p=2$, as does the theory for the reasons discussed above, and is starting to constrain somewhat lower powers (albeit still at low significance).  We cannot quite yet make statistically significant distinctions within this class of models, but this is likely to change in the coming years.  They will be significantly tested by the B-mode observations.  As mentioned above, multifield versions of axion inflation such as N-flation push the tilt to the red of the single-field version of a given model, as analyzed for example in \cite{Liddlemulti, Nflation, Danjie}.  

An additional, more model-dependent signature arises from (i) the oscillating term in (\ref{Vplusosc}) \cite{monodromy},  and from oscillations in masses of moduli and wrapped branes \cite{production}.  
On (i), there are some interesting analyses reported in \cite{oscdata}\cite{Planckpapers}\ (with no detection).   In principle this could affect the power spectrum and non-Gaussianity \cite{resonantNG}\ in a dramatic way, but for most purposes the resonant non-Gaussianity contributes greater signal/noise in the power spectrum as compared to the non-Gaussianity.   The case (ii) is a subject of current research \cite{production}, with a potentially much stronger signal in the non-Gaussianity than in the previous case.  An analysis of the bispectrum (3-point function) underway, and a regime where the non-Gaussianity builds up in higher-point functions, for which the optimal search strategy would go beyond the bispectrum.
Both (i) and (ii) are model-dependent in amplitude and in the evolution of the period $f$ during the process, but it is well worth analyzing this in both in the theory and data as far as we can.

There are many more aspects of axions in cosmology than we have been able to cover here.  As we saw above, in string theory axions arise from higher-dimensional analogues of electromagnetic gauge fields; let me mention that people have considered other uses of gauge fields in inflation such as the recent works \cite{chromo}\cite{gauge}.   

\subsection{Kahler moduli and Higgs/Starobinski-like inflation}     

Some scenarios such as \cite{stringStarobinsky}\ lead to a similar potential for the canonically normalized field as the original Starobinsky model of inflation.  See \cite{inflationreviews}\ for details and caveats, but the basic idea is as follows.  Consider a $D=10$ type IIB string compactification on a Calabi-Yau manifold as in KKLT, but incoporate an $\alpha'$ correction to the Kahler potential, writing
\begin{equation}\label{Kahler}
K=-3 log(\rho+\bar\rho)-\frac{\hat\xi}{{\cal V}}=-2 log {\cal V}-\frac{\hat\xi}{{\cal V}}
\end{equation}
where ${\cal V}$ is the internal volume and $\hat\xi$ is proportional to a topological quantum number which can be sufficiently large to make this correction significant.  Let us also keep track of multiple Kahler moduli $\rho_I$, whose real part contains the $Ith$ 4-cycle volume, with superpotential of the form 
\begin{equation}
W=W_0+\sum A_I e^{-\rho_I}
\end{equation}   
The idea is to seek a solution where many of the four-cycles are small compared to ${\cal V}^{2/3}$.   Then their exponentially suppressed contributions can compete with the power law obtained by expanding (\ref{Kahler}) to stabilize the volume.  The other Kahler moduli remain flat at this order, and provide candidate inflatons. 
They lead to a potential of the form \cite{stringStarobinsky}
\begin{equation}
V(\phi)\simeq V_0\left( 1-\frac{4}{3}e^{-\phi/\sqrt{3}}\right)
\end{equation}
which is similar to that in Starobinsky inflation in terms of the canonically normalized field.        

At a low energy supergravity level, negatively curved field spaces have played an interesting role in connecting to the Staroinbky type model, as in \cite{KalloshLinde}.  Although it is not imposed from top-down considerations, low-energy supergravity is an interesting possibility whose structural constraints and their implications for inflation are very interesting to explore.   Since our focus is on string theory (with or without low energy SUSY), we will refer the reader to that rather extensive literature for details.

\subsection{Gravitationally redshifted D-brane inflation and strings}

A classic example of string-theoretic inflation is the KKLMMT model \cite{KKLMMT}.   By emphasizing the role of Planck-suppressed contributions to the effective theory, this paper played a very important role in enforcing the necessary standard of control on inflationary model-building in string theory.   Being an early example, it is covered in numerous earlier reviews and papers, so here we will be brief.  The basic elements are a D-brane and anti D-brane stretched along our 3+1 dimensions, and living at points in the extra dimensions in a region of strong gravitational redshift, approximately the $AdS_5$ metric
\begin{equation}\label{AdSdS}
ds^2|_{`warped~throat'}\approx  sinh^2\frac{w}{R} ds^2_{dS_{4}}+dw^2
\end{equation}   
where $ds^2_{dS_4}$ is the metric of four-dimensional de Sitter spacetime (i.e. the first approximation to inflation).
This metric applies along a slice of the spacetime going up to some finite value of $w$ where the geometry matches on to a compact manifold \cite{RS, GKP, KKLT}.  
  
The redshift makes it possible for the slow-roll conditions to be satisfied at the level of the classical potential generated by the brane-antibrane interaction energy.  There are, as emphasized in \cite{KKLMMT}, many other contributions to the potential which affect the slow roll parameters at order 1, and these can be packaged in a useful way using the AdS/CFT correspondence to relate them to operator dimensions in a dual field theory description of the warped throat geometry (\ref{AdSdS}), and treated statistically.  As a small-field model without a symmetry protecting the potential, any given realization of the model is somewhat tuned, but there are plausibly many possible realizations.  

The tilt is not determined, as it depends on the Planck-suppressed contributions in any realization of the model, but the tensor to scalar ratio is robustly predicted to be too small for detection; this is a small-field model.  
Because of the warping, a very interesting albeit model-dependent potential signature is cosmic strings from the exit.  The redshifting to low energies makes this viable -- without it such strings would have large enough string tensions that they would already be ruled out  \cite{CSreview}.  Searches for cosmic strings have been made using various probes including \cite{Planckpapers}, with no detection thus far.  

\subsection{DBI Inflation and Equilateral non-Gaussianity}

In this section we will describe a (then-)novel mechanism for inflation that came out of the AdS/CFT correspondence in string theory and our early attempts to generalize it to cosmology.\footnote{This will be the subject of the final part of these lectures.}  The original setup inspired by string theory is similar to that of the previous subsection, but in a different regime  (there are more recent ideas such as \cite{unwinding, TolleyWyman}\ where it may arise in a different way with less extreme parameters).  

Regardless of the particular embedding in string theory, the mechanism helps make very clear how much more general inflation is than the slow-roll case (\ref{SR}).  The requirement for inflation is not (\ref{SR}), but only (\ref{Hslow}), as the following dynamics illustrates explicitly.

Recall our expression for a relativistic particle action (\ref{BI}) which we generalized to the DBI action on branes (\ref{BI}).
As for the particle, this action enforces the fact that the brane cannot move faster than light.  This continues to hold in the presence of a nontrivial potential $V(\phi)$  -- including potentials which are too steep for slow-roll inflation.  The interactions in (\ref{DBI}) slow the field down even on a steep potential.  

Let us evaluate this in the anti-de Sitter geometry
\begin{equation}\label{AdSpoinc}
ds^2=\frac{r^2}{R^2}\left(-dt^2+d{\bf x}^2\right) + \frac{R^2}{r^2}dr^2,
\end{equation}  
putting the brane at a position $r$ that might depend on time $t$.  Here we can set $B=0, f=0$ in (\ref{DBI}), and choose the simplest embedding $\xi^0=t, {\xi^i}={x^i}$.  The metric $G_{MN}$ is given by (\ref{AdSpoinc}).  Plugging all this in, and rewriting $\phi=r/\alpha'$, $\lambda=R^4/{\alpha'}^2$, we obtain the action
\begin{equation}
S=-\int d^4 x \left\{\frac{\phi^4}{\lambda}\sqrt{1-\frac{\lambda\dot\phi^2}{\phi^4}}+\Delta V(\phi)\right\}
\end{equation}  
for a canonically normalized scalar field $\phi$.  The term $\Delta V$ has to do with other contributions to the scalar potential, which depend on the charge and tension of the brane in this noncompact background (\ref{AdSpoinc}).
Regardless of the potential, the field is limited by the speed of light:
\begin{equation}\label{speedlimit}
\frac{\lambda \dot\phi^2}{\phi^4} < 1.
\end{equation}
This makes an infinite difference to the dynamics near $\phi=0$.  If we expand out the square root and consider the metric term alone, the field would roll through the origin $\phi=0$ according to the solution $\phi=\phi_0+v(t-t_0)$, taking a finite time to get to $\phi=0$ from any finite value $\phi_0$.  In contrast, with the interactions in place enforcing (\ref{speedlimit}), it never gets there!

These features persist when we couple this theory to gravity and introduce a potential $V(\phi)$, providing a mechanism for inflation very different from slow roll inflation, one which works for steep potentials.  (This does not imply that it suffers from less tuning than in the slow roll regime, since the rest of the action may require tuning.)  Because of the interactions, the perturbation spectrum is much more non-Gaussian than in slow roll inflation and require a more general analysis \cite{Creminelli}\cite{DBI}\cite{generalsingle}\cite{EFT}\cite{shape}.  Writing $\phi=\phi_0(t)+\delta\phi(t,{\bf x})$ and expanding the square root makes this immediately clear since this brings down inverse powers of the square root.  That is, positive powers of $\gamma=1/\sqrt{1-\lambda\dot\phi_0^2/\phi_0^4}$ multiply interaction terms of order $\delta\phi^3,\delta\phi^4$ and so on.  

The interactions in this theory, expanded in powers of $\lambda\dot\phi^2/\phi^4$, are reminiscent of 
 those in the general discussion above of effective field theory (\ref{Wilsac}).  Except here, the scale $M_*$ is replaced by the field $\phi$ itself.  We do not have time to explain it here, but this lines up very well with the dual description of the system (\ref{AdSpoinc})\ according to the AdS/CFT correspondence \cite{AdSCFT}, where this action arises precisely from integrating out degrees of freedom $\chi$ whose masses are proportional to $\phi$. That is $M_*=M_\chi=\phi$ is indeed a mass threshold in the theory.  

For $\lambda \gg 1$, the effects from integrating out the $\chi$ field dominate over effects of their time-dependent production.  In that sense, trapped inflation \cite{Trapped}\ (discussed above), where the field is slowed down due to dissipation into $\chi$ fields, is a similar mechanism in a different regime of parameters.  These play a role in the scenario \cite{unwinding}, which also has an interesting way of starting the process through explicit bubble nucleation.    
  
The extensive analysis of non-Gaussian shapes using the current CMB data \cite{Planckpapers}\ has provided important constraints on this mechanism and others that produce shapes of non-Gaussianity that have been explicitly searched for.   The strongest constraints apply to
$f_{NL}^{local}$, which is a shape that can only be produced in multifield inflation.   Although a significant portion of the parameter space has been removed, there is still room within the current constraints for substantial non-Gaussianity in the equilateral and orthogonal shapes, among others.  From the point of view of the low-energy effective theory of inflationary perturbations \cite{EFT}, one can quantify the level of constraint required to show that slow roll inflation is favored over other possibilities.  Although all the data is consistent with single-field slow roll, more stringent bounds are needed in order to reach that conclusion.
As a result there is a strong push to analyze large-scale structure in sufficient detail to use it as a probe of non-Gaussianity
 (see e.g. \cite{Snowmass}\ for a recent summary).      
 
\subsection{Additional developments}

In recent years, there has been progress in the understanding of the symmetry structure of certain models of inflation or the ingredients entering into them \cite{KalloshLinde}\cite{KalloshD3}.  For example, the antibrane of KKL(MM)T describes low energy (gravitationally redshifted) supersymmetry breaking in four dimensions, but the appropriate supersymmetric packaging on the gravity side has been mysterious.  This has been addressed in terms of constrained multiplets in \cite{KalloshD3}.  The effective field theory description of the anti-D3-brane was treated in \cite{SBD3}, recovering the picture of \cite{KKLT, otherD3}.        

\subsection{Planck-suppressed operators from hidden sectors}

The limits on certain shapes of non-Gaussianity can provide interesting constraints on theoretical parameters.  One immediate application is to analyze the constraints they imply on Planck-suppressed operators that could couple observed physics to an otherwise hidden sector of additional fields \cite{Plancksuppressed}.  Consider the possibility of such an additional sector of fields.  From our general introduction to effective field theory (\ref{Wilsac}), we would expect couplings between additional fields and the inflaton, suppressed by appropriate energy scales $M_*$, for example
\begin{equation}
\int d^4x \sqrt{-g}\frac{(\partial\phi)^2O_{\Delta}}{M_*^\Delta}
\end{equation}
where $\Delta$ is the dimension the operator $O$.  This includes a mixing term $\dot\phi\dot{\delta\phi}O_\Delta/M_*^\Delta$ between the perturbation $\delta\phi$ of the inflaton and the operator $O$ of the other sector.  In this other sector, there may be significant interactions, including a nontrivial three-point function $\langle OOO\rangle$.  This combined with the mixing interaction induces non-Gaussianity $\langle\zeta\zeta\zeta\rangle$.  When one plugs in the numbers, this leads to constraints on hidden sectors connected via $M_P$-suppressed operators, in the case of high-scale inflation (in general the results depend on the ratio $H/M_*$).  This was just the basic idea; see the papers \cite{Plancksuppressed}\ for a careful treatment.   Recently this was extended to consider the imprint on non-Gaussianities of derivatively coupled massive fields, motivated in part by the spectrum of heavy fields and strings that arise in string theory \cite{juannima}.  

\subsection{Entry and Exit physics}

There have been interesting explorations of the physics and potential for observables coming from the entry into inflation or from the exit phase.  Tunneling into an inflationary trajectory from a metastable vacuum could give very interesting signatures \cite{bubbles}.  Note that this requires a minimal number of e-foldings, whereas  inflation in string theory can naturally produce significant numbers of e-foldings beyond the $\approx 60$ observed without any additional fine-tuning.  (Claims that a small number of e-foldings is preferred are not reliable, as they are based on particular fine-tuned small-field classes of models; nonetheless it is a very interesting possibility.)   Reheating dynamics has brought interesting novelties such as oscillon configurations \cite{oscillons}\ and large-scale non-Gaussianities \cite{Dick}.  We mentioned cosmic strings, another interesting possibilty for observable physics coming out of the exit from inflation, in the context of the model \cite{KKLMMT}\  above \cite{CSreview}.      

\section{What is the framework?}

Let us now switch gears and discuss conceptual questions associated with inflation, and with the late-time accelerating universe \cite{SNDE}\cite{Huterer}.  
The causal structure of de Sitter spacetime is very different from Minkowski spacetime or $AdS$, in that no single observer can collect all the data that appears to exist mathematically in the global geometry.  This question persists even when we include an exit from inflation for any given observer -- because of the structure of the moduli potential that we found above, which runs away toward zero at weak coupling or large radius, there is a nonzero amplitude form a Coleman-de Luccia bubble with that runaway phase inside.  Given this decay, an observer can access more degrees of freedom than in the original de Sitter phase, as previously super-horizon modes come into the horizon; but with inflation persisting eternally in the global sense (going on elsewhere), observer-dependent horizons remain.  

In string theory, there is a well supported conjecture for a complete formulation of physics in Anti de Sitter space (and some other non-cosmological spacetimes) in terms of a dual quantum field theory which is formulated on a spacetime of one less dimension.  This realized an older idea of `holography' that developed out of black hole physics, in which the area of the event horizon behaves like a statistical-mechanical entropy.  The latter idea seems more general than its implementation in $AdS$ and related spacetimes, and we would like to explore its upgrade to inflationary spacetimes, particularly de Sitter.

Our strategy is simply to add ingredients to the `gravity side' of the correspondence, and see what this does to the dual description of the system.  We find nontrivial, but qualitative, agreement between a macroscopic and microscopic approach to this problem, between \cite{dSdS} and \cite{micromanaging}.      

To begin let us introduce the basic ideas behind the AdS/CFT correspondence.  Let us return to the branes we mentioned earlier as stress-energy sources in string theory.  If we introduce a stack of parallel $N_3$ of D3-branes in $D=10$, they source a metric
\begin{equation}\label{threebrane}
ds^2 = \frac{1}{(1+\frac{R^4}{r^4})^{1/2}}(-dt^2+d{\bf x}^2)+(1+\frac{R^4}{r^4})^{1/2}dr^2 + R^2 d\Omega^2
\end{equation}          
where $d\Omega^2$ is the metric on the 5-sphere which surrounds the D3-branes.   The form of the Newtonian potential far from the source is just the generalization of the familiar $1/r$ potential in four dimensions; in general it goes like $1/r^{d_\perp-2}$ where $d_\perp$ is the spatial codimension (the number of spatial directions transverse to the object--here that is 6, so we get 4=6-2).   
These branes are charged: in addition to sourcing a gravitational potential, they also source a five-form flux $F_5=dC_4$ with both electric and magentic components; the quantized internal magnetic flux satisfies $\int_{S^5} F_5=N_3$.        

The redshift factor $G_{00}=1/(1+\frac{R^4}{r^4})^{1/2}$ becomes very small for $r\ll R$.  That is, $r\ll R$ is the low energy regime of this system, where we measure the energy with respect to time $t$.  The metric simplifies in this limit, becoming  the Anti de Sitter metric (\ref{AdSpoinc}) times the $S^5$ metric.  

This $AdS$ solution can be understood as coming from the stabilization of the $S^5$, which occurs via a balance of forces between two terms.  We can think about this in terms of the five-dimensional potential energy as a function of the sphere size and the string coupling, starting from (\ref{Dac})(\ref{Smatter}).  As explained in detail in \cite{TASIlectures}, this produces a potential that depends only on one combination $\eta=e^{\phi_s/3}/L^{4/3}$, where $L=R/\sqrt{\alpha'}$ is the sphere size in units of the string tension.  This potential is of the form
\begin{equation}
V=M_5^5\left(-\eta^4+N_3^2 \eta^{10}\right)
\end{equation}
The first, negative term comes from the positive curvature of the $S^5$ in the $D=10$ Einstein term $\int\sqrt{-G}R$.  The second term comes from the flux term $F_5^2$, which pushes against the contraction of the sphere since that causes large energy density.  In more detail, the curvature term goes like $-(L^5/g_s^2)\times(1/L^2)$ times a Weyl rescaling factor to go to Einstein frame, and the flux goes like $L^5 \times N_3^2/L^{10}$ times the conversion factor.

There is another description of low energy physics in this system:  that of the open string theory on the D-branes, which make up a specific quantum field theory, the $N=4$ supersymmetric Yang-Mills theory with gauge group $U(N_3)$.  This is a distant cousin of quantum chromodynamics, a $U(3)$ Yang-Mills theory.  The conjecture is that these two low energy descriptions are equivalent \cite{AdSCFT}.  

We will want to draw from two generalizations of this structure which give additional examples of the duality.  The first is to consider a stack of D3-branes at the tip of a cone whose base $X$ is a positively curved Einstein space (meaning $R_{ij}=const\times g_{ij}$).  This also gives $AdS$ solutions dual to specific field theories, as in \cite{orbifoldCFT}\ and many generalizations.  It is useful to describe the cone (which is simply locally flat space), or the original 6-dimensional flat space in the original example above, 
\begin{equation}\label{cone}
ds^2=dr^2+R(r)^2 ds^2_{X}=ds^2=dr^2+ r^2 ds^2_{X}
\end{equation}
as the solution to a radial version of the Friedmann equation
\begin{equation}\label{radialAdS}
\left(\frac{R'}{R}\right)^2=\frac{1}{R^2} \Rightarrow R(r)=r
\end{equation}
where the role of the scale factor is played here by $R(r)$.  
Here the $1/R^2$ on the right hand side comes from the curvature of the base (e.g. $S^5$ in the original example above).  To belabor the obvious, the solution $R(r)$ goes off to infinity as $r\to\infty$.      
The reason for making these comments is that when we `uplift' to de Sitter spacetime it will be useful to see how (\ref{radialAdS}) is modified.   

Let us now analyze what happens when we add contributions to the potential to produce metastable de Sitter instead of anti de Sitter.  Recall that the stabilization mechanism involved two terms:  the sphere (or more general Einstein space) curvature, and the $N_3$ units of flux sourced by the D3-branes.  Placing the D3-branes at the tip of the cone causes the geometry to become $AdS_5\times X$ in the small-$r$ `near horizon' region.  As we discussed above, the stabilization of moduli to metastable de Sitter spacetime requires a three-term structure (at least), a potential of the schematic form
\begin{equation}\label{Vabc}
V=a\frac{1}{R^{n_a}}-b\frac{1}{R^{n_b}}+c\frac{1}{R^{n_c}}
\end{equation}  
where $n_c>n_b>n_a$ and $a,b,c>0$.  Here in general the coefficients $a,b$, and $c$ will depend on many scalar field moduli, and in each direction the potential must stabilize these or at least produce accelerated expansion.  This has been done in an explicit uplift of a different version of  $AdS/CFT$ \cite{micromanaging}, and it is rather complicated but does have this basic structure, with the final term proportional to $c$ coming from fluxes with indices along the internal sphere
($F_5$ flux in the above example).  We will be interested in the qualitative features here.  

We could consider for example the case that the first, positive term proportional to $a$ comes from changing the internal space $X$ from being positively curved to being negatively curved.  This is possible to do, with sources of stress-energy which have a known field-theoretic interpretation \cite{dualpurpose}\cite{micromanaging}\cite{FRW}\ in terms of magnetic matter.  Another possibility is to introduce branes which wrap $X$.   We need a negative term which can come from the exotic objects we discussed above, `orientifolds'.  

Given those elements, we can now see what happens to the radial Friedmann equation (\ref{radialAdS}) upon such an uplift of $AdS/CFT$.  It becomes
\begin{equation}
\left(\frac{R'}{R}\right)^2=-\frac{1}{R^{n_a}}+\frac{1}{R^{n_b}} 
\end{equation}    
with the flux term left out as before because it corresponds to the D-branes we placed at the origin of the cone.
With $n_a<n_b$, this equation has a solution in which the $R(r)$ starts small, grows to a finite maximal value, and then shrinks again.  The cone has become like a rugby ball (or American football) in shape, with two tips and a finite maximum of $R(r)$ in between.     

\begin{figure}[h!]
\begin{center}
\includegraphics[width=12cm]{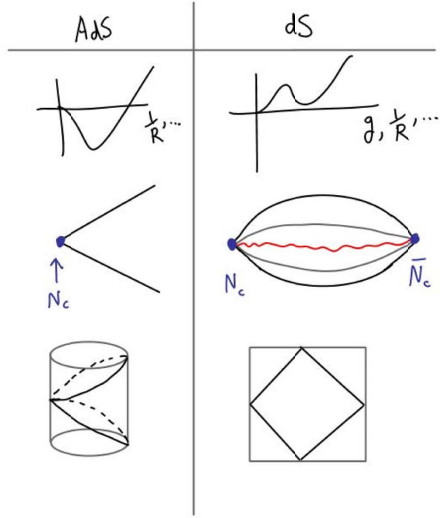}
\caption{\label{fig:dSdS} \small The structure of the string landscape -- specifically its metastability -- leads to a brane construction for de Sitter spacetime which nontrivially agrees with the macroscopic structure of de Sitter as a two-throated warped compactification.  This would not have happened if string theory admitted a hard cosmological constant parameter.  The left panel depicts AdS/CFT dual pairs arising from compactification on a positively curved Einstein space.  This leads to the leading negative term in its potential, and appears as the base of the cone in the brane construction, D-branes probing the tip of a cone whose size satisfies a radial Friedmann equation.  Trading the branes for flux and geometry, the resulting AdS solution has a warp factor that extends to the deep UV.  The right panel depicts the effect on all this of uplifting to a metastable de Sitter solution.  The leading term in the potential is now positive, e.g. coming from negative curvature.  This combined with the second, negative term in the potential (e.g. from orientifolds) leads to a de Sitter analogue of the brane construction whose radial Friedmann equation implies that the size grows to a finite maximum before contracting.   This fits perfectly with the dS/dS slicing of macroscopic de Sitter spacetime, which exhibits two warped throats cutoff at a finite UV scale.  This structure occurs also in the computation of observables in the dS/CFT approach, which requires integrating over the $d-1$ dimensional metric.  All this suggests that the matter sector in the appropriate holographic dual to de Sitter is a theory that is not UV complete (perhaps analogous to QED), in contrast to gauge/gravity duals with asymptotic boundaries.}
\end{center}
\end{figure}

Since adding $N$ units of flux produces a metastable dS solution, we can obtain that solution by introducing $N$ branes at one tip and $N$ anti-branes at the other; these have an equivalent description in terms of the flux.  This introduces {\it two} low energy sectors. 

This lines up beautifully with the geometry of de Sitter spacetime.  The latter has a metric (which is not global, but covers more than an observer patch) 
\begin{equation}\label{dSdS}
ds^2_{dS_{d}}=sin^2(\frac{w}{L})ds^2_{dS_{d-1}}+dw^2 
\end{equation}
This exhbits a gravitational redshift which mirrors the structure we just derived from the brane construction:  namely, $g_{00}$ starts at zero at $w=0$ where $sin(w/L)=0$, it rises to a finite maximum at $w=\pi L/2$, and decreases again to a second zero at $w=\pi L$.  That is, there are two low energy regions.  This is what we just found above:  two low energy sectors from the two tips.  

This agreement is rather striking, even though it is only qualitative.  Note that it would not have occurred if string theory came with a hard cosmological constant:  the vanishing of the potential in large-radius limits, encoded in the structure (\ref{Vabc}), plays a crucial role here.   

The analogous metric on $AdS$ spacetime is (\ref{AdSdS}), with the coordinate $w$ and the redshift factor $g_{00}$ going all the way up to infinity.  This regime is the deep ultraviolet region of the field theory, described by local operators.  In warped compactifications \cite{RS}\cite{GKP}\cite{KKLT}\cite{KKLMMT}, or as we have just seen in de Sitter spacetime itself \cite{dSdS}, the redshift factor goes down to zero in the infrared, but not all the way up to infinity.  One important consequence of this is that the system has dynamical $d-1$ dimensional gravity.  The dual description is a pair of low-energy quantum field theories, coupled via $d-1$ dimensional gravity.  

The dual field theories need not be ultraviolet complete \cite{Dusan}; they only need a good low energy regime, as with quantum electrodynamics.   
In fact, it is a tautology that a theory that is dual to $AdS$ cannot be dual to $dS$, and similarly for their respective generalizations to RG flows or rolling scalar FRW solutions.  The fact that almost all string backgrounds have positive potential energy matches well with the fact that almost all gauge theories are not asymptotically free or conformal:  there are infinite sequences of irrelevant operators that can be present.  

Since three-dimensional gravity is much simpler than four-dimensional gravity -- and since the low energy region near the horizon is dualized in terms of a non-gravitational theory -- this represents some progress, albeit not a complete formulation of the theory.   One can also dimensionally reduce further to obtain two-dimensional, Liouville gravity \cite{dSdS} (as was also found in a different framework \cite{FRWCFT}).  This seems to me to contain the right physics -- it reflects the fluctuating nature of cosmological solutions at finite time and the finite Gibbons-Hawking entropy of de Sitter spacetime.  Other approaches to the problem, including the conjectured dS/CFT correspondence \cite{dSCFT}, have the same feature:  in order to calculate observables, one must make sense of integration over $d-1$ dimensional metrics, i.e. dynamical $d-1$ dimensional gravity.  In fact, those calculations also involve two matter sectors coupled through the $d-1$ dimensional metric,
\begin{equation}
\langle {\cal O}\rangle \sim \int D g^{(d-1)}_{\mu\nu} \Psi^\dagger(g^{(d-1)}_{\mu\nu}){\cal O}\Psi(g^{(d-1)}_{\mu\nu})
\end{equation}
where $\Psi$ is computed by a CFT partition function with the latter living on a space with metric $g^{(d-1)}_{\mu\nu}$, and similarly for $\Psi^\dagger$ which would be computed by an isormophic CFT. 
Thus, there are again two matter sectors coupled to gravity in one less dimension, similarly to what we derived above using the dS/dS slicing.
Perhaps this is not a coincidence.\footnote{We thank L. Susskind for this observation.}

In these systems there is actually an important simplification that arises in the far future, something that also follows from the structure of the potential \cite{FRW, HarlowSusskind}.  Because the de Sitter solutions are only metatable, they eventually decay to a rolling scalar FRW solution with decelerating expansion.  The causal structure within an observer patch becomes more like Minkowski space in the future.  The gravitational entropy bound \cite{Boussobound}\ goes off to infinity.  The analogue of the warped metric  (\ref{dSdS}) has the property that the inferred $d-1$ dimensional Newton constant goes to zero at late times -- the $d-1$ dimensional gravity eventually decouples!  In fact, this is true also for accelerating solutions with $w>-1$ as in \cite{SCDE}, even though their causal structure is similar to de Sitter \cite{HellSuss}.

These features strike me as very promising, but there is much more to do to flesh out and test these ideas.  Simpler solutions will help, and there are many ideas to pursue for reducing the list of sources required to generate inflation (e.g. \cite{SCDE, Danielsson}).  There is another general approach which we do not have time to cover here: one can obtain clues about the dual description by trading fluxes for branes and analyzing the theory as it moves out along the resulting space of scalar fields.  

In general, the structure of spacetime dependent field theory and string theory is an extremely fruitful area for further research.  The bulk of research in string theory thus far is on special, highly symmetric solutions in which
one can compute certain quantities very elegantly even at strong coupling.\footnote{One should never confuse the statistics of string theory papers with the statistics of string theory backgrounds.}  However, even from the point of view of mathematical physics, the more general setting of curved target spaces with nontrivial evolution are extremely interesting -- they exhibit beautiful generalizations of string dualities involving their topology and geometry.  In any case, the observational discovery of the accelerating universe and the evidence for inflation in the primordial perturbations provide ample motivation for further work toward a complete framework.  

\section*{Acknowledgements}

I am grateful to numerous collaborators and other colleagues for many illuminating discussions in this exciting area of physics.
I would like to thank the organizers and participants for this very stimulating school, including some spectacular non-slow-roll bike rides in the local mountains.  I am also grateful to the participants and organizers of the 2013 Les Houches school on Post-Planck Cosmology, the 2013 ICTP Spring School, the 2011 PiTP school Frontiers of Physics in Cosmology at the IAS, and the 2010 La Plata, Argentina school on high-energy physics where earlier versions of these lectures were given.   This work was supported in part by the National Science Foundation under the grant
PHY-0756174, and by the Department of Energy under contract DE-AC03-76SF00515.

\end{document}